\pdfoutput=1

\documentclass{article}
\usepackage{anysize, color}
\marginsize{0.7in}{0.7in}{0.7in}{0.8in}

\usepackage{color}
\usepackage{hyperref}
\usepackage{graphicx}
\usepackage{subcaption}

\def\ket#1{\mathinner{|{#1}\rangle}}
\def\rCZ{{\rm CZ}}
\def\rZ{{\rm Z}}
\def\rX{{\rm X}}
\def\rY{{\rm Y}}
\def\rI{{\rm I}}
\def\rP{{\rm P}}
\newcommand{\Xsqr}{X$^{\frac{1}{2}}$}
\newcommand{\Ysqr}{Y$^{\frac{1}{2}}$}

\begin{document}

\title{Quantum Supremacy Is Both Closer and Farther \\
        than It Appears}
\author{Igor L. Markov$^1$, Aneeqa Fatima$^1$, Sergei V. Isakov$^2$, and Sergio Boixo$^3$  \\
            $^1$ University of Michigan,  2260 Hayward St, Ann Arbor, MI 48109 \\
            $^2$ Google Inc., 8002 Z\"urich, Switzerland\\
            $^3$ Google Inc., Venice, CA, 90291
            }
\maketitle

\abstract{
As quantum computers improve in the number of qubits and fidelity, the question of when they surpass
state-of-the-art classical computation for a well-defined computational task is attracting much attention.
The leading candidate task for this 
milestone entails sampling from the output distribution defined by a random quantum circuit. We develop
a massively-parallel simulation tool Rollright that does not require inter-process communication (IPC) or proprietary hardware. We also develop two ways to trade circuit fidelity for computational speedups, so as to match the fidelity of a given quantum computer --- a task previously thought impossible. We report massive speedups for the sampling task over prior software from Microsoft, IBM, Alibaba and Google, as well as supercomputer and GPU-based simulations. By using publicly available Google Cloud Computing, we price such simulations and enable comparisons by total cost across hardware platforms. We simulate approximate sampling from the output of a circuit with $7 \times 8$ qubits and depth $1+40+1$ by producing one million bitstring probabilities with fidelity 0.5\%, at an estimated cost of \$35184. The simulation costs scale linearly with fidelity, and using this scaling 
we estimate that extending circuit depth to $1+48+1$ increases costs to one million dollars.
Scaling the simulation to 10M bitstring probabilities needed for sampling 1M bitstrings helps comparing simulation to quantum computers. We describe refinements in benchmarks that slow down leading simulators, halving the circuit depth that can be simulated within the same time.
 }

\section{Introduction}

 The promise of quantum computers inspired the pursuit of {\em quantum computational supremacy}, i.e.,
 using quantum computers to solve some task that is prohibitively hard for conventional computers \cite{aaronson2011computational,bremner2015average,Preskill,bremner16,GoogleSuprem,NatureSupremacy,aaronson2016complexity,neill_blueprint_2017,bouland_quantum_2018}.
A key benchmark has emerged in terms of random, universal quantum circuits proposed by Google \cite{GoogleSuprem},
and this inspired rapid progress in algorithms for simulating quantum computers~\cite{de_raedt_massively_2007,GoogleSuprem,PB,IBM,GoogleBucket,64q,Taihu,chen_classical_2018,de_raedt_massively_2018}. Simulating sampling or performing cross-entropy benchmarking~\cite{GoogleSuprem} requires calculating a large number of output probabilities requested at random. We focus on a method that essentially evolves the quantum wave function and outputs a large number of probabilities with small additional cost~\cite{aaronson2016complexity,64q}. We show that two recent quantum simulations that required supercomputers with massive amounts of memory \cite{PB,IBM} can now be performed with modest amounts of space and time using publicly accessible cloud computing.
To match the accuracy of results in a given quantum computer with arbitrary quantum gates,
we developed an approximate simulation whose runtime scales linearly with circuit fidelity (a measure of 
how much quantum information is preserved).  In contrast to the recent use of the world's most powerful supercomputer (over 100 Pflops, 131K nodes) to simulate Google circuits~\cite{Taihu}, we add a new dimension to quantum-versus-classical comparisons and report monetary cost of our simulations in terms of cloud-computing resources. This metric captures rapid progress in quantum-circuit simulation since 2016. On the other hand, we show that
\begin{itemize}
\item Small changes to common quantum-supremacy circuits make them considerably harder to simulate.
\item The use of more sophisticated quantum gates substantially handicaps leading simulation methods.
\end{itemize}

In 2017 and 2018, IBM, Intel and Google announced new quantum computing chips that implement 50 and 72 qubits \cite{ibm_2017,intel_18,google_2018}, but no empirical results for these chips have yet been announced as the chips are being improved. These chips will have to compete with simulation software running on conventional computers. The most straightforward approach --- Schr\"odinger-style simulation that maintains and modifies the wave-function --- requires $2^n$-sized memory for $n$ qubits~\cite{de_raedt_massively_2007}. Therefore, it is possible to simulate 30-qubit circuits on a laptop, and 35-qubit circuits on a mid-range server. Further scaling would seem to require supercomputers, and a 2016 simulation \cite{PB} used 0.5 PB RAM on a Cori II supercomputer to simulate a depth-26 45-qubit circuit proposed by Google for quantum supremacy experiments \cite{GoogleSuprem}. These universal circuits were designed to run on planar $\sqrt n  \times \sqrt n$ qubit-array architectures \cite{GoogleSuprem} and evade easy simulation.  They start with a Hadamard (H) gate on every qubit and arrange nearest-neighbor controlled-Z (CZ) gates in a repeated pattern, so that every possible CZ gate appears once every eight cycles. One-qubit \Xsqr, \Ysqr and T gates are randomly interspersed between CZ gates so as to prevent cancellations and induce chaotic quantum dynamics~\cite{GoogleSuprem}, as illustrated in \cite[Figure 1]{PB}. Each circuit ends with a measurement on every qubit. On a quantum computer this measurement produces an unpredictable bitstring and thus samples the distribution determined by the specific quantum circuit. Competing simulation techniques must calculate a large enough number of amplitudes (each corresponding to some bitstring) to simulate sampling from the output distribution \cite{neville_classical_2017,GoogleSuprem,GoogleBucket}.

Common sense suggests that circuits with more than 49 qubits would require too much memory to simulate in practice. To the contrary, alternative methods related to
Feynman paths~\cite{bernstein1997quantum}, tensor network contractions~\cite{markov_simulating_2008,GoogleBucket}, and similar
approaches~\cite{aaronson2016complexity,64q} can trade space for time complexity. Google researchers have shown that for circuits with low depth, one can quickly find any one amplitude on a single computer for over 100 qubits using a variable elimination algorithm~\cite{GoogleBucket}, and have illustrated the statistical distribution of the output of a circuit with $7 \times 8$ qubits and depth 30 by computing $2\times 10^5$ probabilities using multiple computers. 
In April 2018, a simulation on the world's largest supercomputer, Sunway TaihuLight \cite{Taihu}, computed $2^{46}$ amplitudes for a circuit with $7 \times 7$ qubits and depth $1+39$. Our notation for depth (introduced here) explicitly denotes layers of Hadamard gates with ``$1+$'' at the beginning of the circuit (clock cycle 0) and ``+1'' at the end. The remaining 39 layers (clock cycles) include CZ gates and other, one-qubit, gates. A related computation produced a single amplitude for a circuit of depth 55. The depth attained in these simulations includes 8 cycles gained by exploiting an unfortunate design choice --- sequences of diagonal gates \verb+CZ - T - CZ+, --- at the beginning of the circuit, specific to the circuits in Ref.~\cite{GoogleSuprem} and avoided in revised benchmarks~\cite{new_benchmarks} as explained in Section
\ref{sec:bench}.

All simulations discussed so far seek exact output amplitudes, aside from negligible numerical errors. However, the computational task of interest is to {\it approximately} sample from the bitstring distribution defined by a random quantum circuit, and near-term quantum computers incur significant, unavoidable errors because gate errors accumulate exponentially~\cite{Preskill}. In particular, for Google quantum supremacy circuits~\cite{GoogleSuprem,new_benchmarks} with $7\times7$ qubits and depth $1+40+1$, a reasonable goal is to achieve a 0.005 circuit fidelity, consistent with a two-qubit gate fidelity of 0.995, one-qubit gate fidelity of 0.999, initialization fidelity of 0.998 and measurement fidelity of 0.99~\cite{barends_superconducting_2014,barends_digitized_2015,kelly_state_2015,GoogleSuprem} (in this example, depth $1+40+1$ entails initial and final Hadamard gates on each qubit). The same circuit fidelity at depth $1+48+1$ would require a two-qubit gate fidelity of 0.996. In our simulations, we configure circuit fidelity values $f=0.01$ and $f=0.005$, and show how to revise resource estimates to any given $f\in (0, 1]$, using linear scaling that we prove. Significantly, rather than simulate individual gates to fixed fidelity, we control simulation fidelity for the entire circuit.
 
Extending recent progress in quantum circuit simulation algorithms, coauthors from the University of 
Michigan implemented Rollright --- a massively-parallel simulation tool that does not require inter-process communication (IPC) or proprietary hardware. Rollright can run in commercial cloud-based {\em preemptible virtual machines} (VMs).\footnote{A preemptible VM is cheaper than an on-demand VM because it does not guarantee real-time execution. A process may be terminated (preempted) by a higher-priority task at any time, but can then be restarted later. Pricing for preemptible VMs in a commercial cloud varies for different hardware configurations and may fluctuate hourly with demand for computational resources.} Table \ref{tab:supercomp} summarizes advantages of this approach over supercomputer simulations \cite{PB,IBM,Taihu}.
 We report the following results:

\begin{itemize}
 \item Using a MacBook Pro laptop with 16GiB RAM, Rollright can simulate circuits with $5 \times 5$ qubits
         to any depth, with $16\times$ and $10\times$ speedups over the latest versions of simulators from
         Microsoft QDK (Appendix~\ref{app:QDK}) and IBM QISKit / QASM (Appendix~\ref{app:QASM}),
         respectively.  Rollright uses $3.27\times$ and $1.7\times$ less memory respectively and can also
         simulate $6\times5$-qubit circuits of any depth on the MacBook. On a single 72-thread server,
         Rollright in its depth-limited mode simulates a depth-28 circuit with $6\times5$ qubits $31\times$
         and $16\times$ faster than Microsoft and IBM simulators, respectively. Rollright uses $886\times$
         and $590\times$ less memory for this simulation.\footnote{These results were presented at Microsoft and IBM.}
         Running QISKit on GPU-based servers through the IBM Q service is $12\times$ slower than Rollright,
         which uses no GPUs. The speedup increases to $877\times$ for a 32-qubit circuit.
\item Using a single {\tt n1-highcpu-96 server}
       from Google Cloud, we calculate one million probabilities (requested at random) per simulation, enough to simulate sampling~\cite{neville_classical_2017,GoogleBucket} for circuits used in two recent supercomputer simulations:
       \begin{itemize}
       \item For a circuit with $6 \times 7$ qubits and depth $1+25$~\cite{GoogleSuprem},
       our simulation took 4.7 hours, with estimated cost \$3.34 in the cloud.
       This is significantly cheaper and more accessible than simulations in Ref.~\cite{GoogleSuprem},
       but we recall
       that those simulations study the convergence to the Porter-Thomas distribution as a function of depth
       with exponential precision.
       \item For a circuit with $9 \times 5$ qubits and depth $1+25$~\cite{PB} our simulation took 20 minutes,
       with estimated cost \$0.24 in the cloud.
       We use 17.4GiB peak memory where Ref.~\cite{PB} used 0.5PB --- a 28600 times
       improvement.\footnote{Our peak memory usage can be improved or traded for runtime at the same
       cost point.}  Notably, for methods that compute one amplitude at a time,
       $9 \times 5$ circuits are not substantially harder than 
       $5 \times 5$ circuits~\cite{GoogleBucket}.
   \end{itemize}
   These results are reported only to compare our simulation to prior work. We also calculate one million probabilities for the circuit \verb+inst_7x6_26_0+~\cite{new_benchmarks} with $7 \times 6$ qubits and depth $1+25+1$. This refined benchmark is substantially harder to simulate for some methods (see Section~\ref{sec:bench}). The runtime was 6 hours (as apposed to 4.7 hours for the previous circuit of similar size), corresponding to a cost of \$4.32.

 \item We propose two complementary new approaches to approximate simulation of quantum circuits
         with arbitrary gate libraries (Sections~\ref{sec:approx} and \ref{sec:sampling}):
         \begin{itemize}
         \item When simulating a quantum circuit, we obtain arbitrary requested sets of output
         amplitudes with prescribed fidelity.  Our ability to produce
         millions of amplitudes in one run implies massive speed-ups over prior works that
         produce one amplitude at a time~\cite{GoogleBucket,chen_classical_2018}.
         \item Given a set of output probabilities, drawn at random, we produce a bitstring
         with a tunable statistical error. Ten randomly chosen probabilities suffice
         to produce one bitstring sample with negligible error. 
         \end{itemize}
         Numerically, we compute one million probabilities with fidelity 0.5\%  for the
         \verb+inst_7x7_41_0+ circuit~\cite{new_benchmarks} with
         $7 \times 7$ qubits, depth $1+40+1$ and controlled-Z (CZ) two-qubit gates. For this,
         we use 625 {\tt n1-highcpu-32} servers from Google Cloud, with estimated cost \$8734.
         We study three types of scaling for resource requirements
         \begin{itemize}
         \item Linear scaling with fidelity $f$.
         \item Nonlinear scaling with the number of amplitudes (Table \ref{tab:10e8}).
         \item Exponential scaling with circuit depth. 
         Extending circuit depth to $1+48+1$ increases cost estimates to one million US dollars. 
         \end{itemize}
\item We produce as many bitstring samples as required
         for cross-entropy benchmarking with the fidelities chosen, enabling cost and runtime comparisons between simulators
         and quantum computers.
\item We propose an interactive protocol to validate the results of quantum-supremacy
      simulations without knowing the correct answer (see Appendix \ref{app:validation}).
      The protocol uses ($i$) approximate simulation with prescribed fidelity, and
      ($ii$) inner-product estimation.
\end{itemize}

Our simulation produces a large number of requested amplitudes at once (cf. \cite{GoogleBucket,chen_classical_2018}) and is not specialized to individual circuits (cf. \cite{Taihu,64q,chen_classical_2018}). In Section~\ref{sec:bench}, we summarize weaknesses found in quantum supremacy circuits \cite{GoogleSuprem} simulated in prior work and introduce new benchmarks that are harder to simulate. Section~\ref{sec:2q} shows that replacing CZ gates with iSWAP gates makes simulation much more difficult yet.

\begin{table}[t]
\begin{center}
\begin{tabular}{|l|l|l|}
\hline
\sc
 & \sc Supercomputer simulations & \sc No-IPC distributed simulations \\
\hline
\bf Total memory and CPUs  & Great, but for short time intervals & Modest, but available for longer time \\
\hline
\bf Researchers' access to HW  & Limited: must adapt to existing    & Easy: cloud services and universities  \\
                                               &  HW config and availability in time   & offer many HW configurations     \\
\hline
\bf Time-space tradeoffs & System- and algorithm-dependent                         & Straightforward (iso total core-hours)   \\
\hline
\bf Sys-specific programming & Extensive (NUMA, data transfers) & None               \\
\hline
\bf Memory buffers      & Needed for data transfers \cite{PB,IBM}  & Not needed                 \\
\hline
\bf SW reuse on diverse HW & Computational kernels reuse & Full SW reuse between systems    \\
\hline
\bf Job scheduling       & Synchronization needed     &  No synchronization needed                       \\
\hline
\bf Communication time-outs & May force a restart    & Easy to tolerate by restarting          \\
\bf and job failures       &   of the entire simulation   & failed jobs only \\
\hline
\bf Preemptible VMs in cloud    &   Incompatible            & Straightforward to use                  \\
\hline
\bf Scalability              & Depends on communication & Linear, unaffected by communication \\
\hline
\bf Performance estimation &   System-specific            & Straightforward                               \\
\hline
\bf Simulation costs      & Significant, hard to quantify & Relatively low, easy to forecast,  but\\
                                 &                                         & may fluctuate with cloud demand \\
\hline
\end{tabular}
\parbox{15cm}{
\caption{\label{tab:supercomp} Comparing simulations of quantum circuits on ($i$) supercomputers \cite{GoogleSuprem,PB,IBM,Taihu} that rely on fast interconnect to ($ii$) simulations on distributed clusters \cite{64q} with no interprocess communication (IPC). 
Differences in the sophistication and performance of CPUs may be reflected in the total cost of simulation. Available memory and its speed may also affect total costs and may encourage different types of simulation algorithms for supercomputer simulations.
}}
\end{center}
\vspace{-4mm}
\end{table}

\section{Our quantum-circuit simulation framework}
\label{sec:our}
Our algorithms can be characterized as Schr\"odinger-Feynman hybrids~\cite{aaronson2016complexity,64q}. In the context of nearest-neighbor quantum architectures, we partition a given qubit layout into blocks. We then decompose quantum gates acting across the partition into sums of separable terms, such that, for each term, each block can be simulated independently and the results can be added up. For example, using a pair of half-sized qubit blocks instead of representing a full wave function reduces memory requirements for $k$ qubits from $2^k$ to $2^{k/2 + 1}$, but introduces a dependency on the number of decomposed gates. For CZ gates the decomposition has two terms\footnote{A two-qubit gate can be decomposed into at most four tensor products, e.g., with one-side operator basis $\rP_{0}$, $\rP_{1}$, $|0\rangle \langle 1|$, $|1\rangle\langle 0|$.}
\begin{equation}
   \rCZ = \mathrm{diag}(1,1,1,-1) = \left(
   \begin{array}{cc}
     1 & 0 \\
     0 & 0 \\
   \end{array}
   \right) \otimes
   \left(
   \begin{array}{cc}
     1 & 0 \\
     0 & 1 \\
   \end{array}
   \right)
   +
   \left(
   \begin{array}{cc}
     0 & 0 \\
     0 & 1 \\
   \end{array}
   \right)\otimes
   \left(
   \begin{array}{cc}
     1 & 0 \\
     0 & -1 \\
   \end{array}
   \right)
   = \rP_{0}\otimes \rI + \rP_{1} \otimes \rZ\;.
\label{eq:CZ}
\end{equation}
Thus, applying each {\em cross-block} CZ gate (xCZ) to a tensor term produces two tensor product terms, doubling runtime.\footnote{Moreover, new pairs of tensor terms are orthogonal because their first components (produced by P$_0$ and P$_1$) are. While unitary operators preserve their orthogonality, projections can increase their cosine similarity.} In comparison, traditional Feynman-style path summation~\cite{bernstein1997quantum,GoogleSuprem} uses very small amounts of memory, but doubles its runtime on every (branching) gate, resulting in much longer
runtime and not being able to use available memory fully. Our simulator combines highly-optimized Schr\"odinger-style simulation within each qubit block and simulates xCZ gates with Feynman-style path summation, to limit memory use. Unlike in Feynman-style simulation, runtime scales with the number of
xCZ gates, which is very limited in planar qubit-array architectures with nearest-neighbor gates \cite{aaronson2016complexity}. Unlike traditional Schr\"odinger-style simulation, the resulting algorithms are depth-limited, and supercomputer simulations may hold some advantage for very deep circuits. However, near-term quantum computers rely on noisy gates \cite{Preskill} that also limit circuit depth.

\ \\
\noindent {\bf Our Schr\"odinger-style simulation} includes
optimizations that help with arbitrary gate libraries and some
that help with small gate libraries, while others target the Clifford+T and related gate
 libraries\footnote{Clifford+T and related gate libraries are common because they promise compatibility with
quantum error-correction, they are used in Google and IBM quantum computers.} These optimizations
equally apply when using supercomputers, laptops and other hardware.
\begin{itemize}
\item Clustering gates of a kind (with reordering), rather than gates acting on the same qubits as in prior simulations. For example, in Google quantum supremacy circuits \cite{GoogleSuprem,new_benchmarks}, we collect separate clusters of CZ, T, \Xsqr and \Ysqr gates.\footnote{This technique might not help in circuits that use a large number of different gate types.}
\item Compact encoding of gate positions in each cluster into bitmasks, so that efficient bitwise operations (parity, population count, as well as counts of trailing and leading zero bits) and mod-8 arithmetics (for T gates) can apply all diagonal gates in a single pass over the wave-function or its slices.\footnote{Some of these techniques exploit specific gate types, and our simulator may run $2-3\times$ slower on a different gate library.}
\item Cache-efficient algorithms to simulate large gate clusters. Instead of accessing
an entire wave-function, we apply gate clusters on amplitude slices sized
to fit in the L2 cache.\footnote{This technique does not depend on specific gate types.}
\end{itemize}
Compared to the state of the art described in \cite{PB}, we can eliminate floating-point multiplications when simulating Google circuits (and only use them for convenience when simulating T gates). The number of floating-point additions is reduced by a factor of three, while fully benefiting from vectorized instruction extensions (AVX2) in modern CPUs. This puts emphasis on cache performance and
limits thread scalability. By streamlining memory accesses, we improve single-thread performance as well
as scaling to multiple threads and multiple processes using the same memory bank. However, the impact of subsequent improvements is even greater.

\ \\
\noindent {\bf For Feynman-style path summation}, our simulator optimizes performance in several ways.
When simulating a circuit with $x>0$ xCZ gates, we use the technique suggested in \cite{aaronson2016complexity} and demonstrated in \cite{64q} where each process receives a bitstring that encodes a unique path, then save the simulation results in a file for subsequent summation.\footnote{Our simulator was implemented before \cite{64q} was described publicly.} This is convenient because no interprocess communication is required, but many processes simulate the same path prefixes. We therefore subdivide $x=x_p + x_b$ and use only $x_p$-bit prefixes to spawn separate processes. After the first $x_p$ bits, each process checkpoints its simulation state (doubling its memory usage once) and simulates each of $2^{x_b}$ remaining paths starting from this checkpoint. A second optimization for Feynman-style path summation is supported by the Schr\"odinger-style simulator algorithm: we noticed that the use of projection operators in gate decompositions such as Equation (\ref{eq:CZ}) leads to a large number of zeros in block-local wave-functions. Our simulator skips blocks of zero amplitudes.

\ \\
\noindent {\bf Final-state amplitudes} are calculated for any requested subset of indices, unlike simulators in
\cite{IBM,Taihu,chen_classical_2018}. This is necessary to simulate sampling~\cite{neville_classical_2017} and to perform cross-entropy benchmarking~\cite{GoogleSuprem}. Subsets of millions of amplitudes are calculated almost as quickly as single amplitudes (c.f.~\cite{GoogleBucket}). However, if a very large subset is requested, such calculations may dominate the cost of ``easy'' simulations. 
Contributions to requested amplitudes are accumulated over simulation paths. When used in its {\em exact mode}, the simulator calculates each amplitude with a negligibly small numerical error.\footnote{As a check, we calculate the norm of the state vector using higher-precision accumulator variables and underflow mitigation. Comparing the result to 1.0, we observe a very close match in all cases when such a match is expected.}

\section{Trading off simulation fidelity for computational resources}
\label{sec:approx}
We are interested in performing the following computational task proposed for a near-term quantum supremacy demonstration: approximately sampling from the output distribution defined by a random quantum circuit~\cite{GoogleSuprem,aaronson2016complexity}. Quantum computers experience
errors in qubit initialization, gates and measurements. This suggests speeding up the simulation by reducing the accuracy of results to the level attained by quantum computers. We are not aware of such attempts for simulating quantum supremacy circuits, but the work in \cite{bravyi_improved_2016, bravyi_2018} developed approximate simulation of circuits dominated by Clifford gates.

\ \\
\noindent {\bf Our approach to approximate simulation}
can handle arbitrary quantum gates, and our fidelity scaling is very different from error scaling in Refs.~\cite{bravyi_improved_2016, bravyi_2018}, as shown below. Equation (\ref{eq:CZ}) shows that each xCZ results in two tensor-product terms (up to four terms for other gates). Simulating $x$ such xCZ gates (among other gates) produces $2^x$ simulation paths, corresponding to all possible choices of either $\rP_0 \otimes I$ or $\rP_1 \otimes \rZ$ for each xCZ, see Equation (\ref{eq:CZ}) . Therefore we can write the output of the simulation as
\begin{equation}
 |\psi\rangle = \sum_{j=0}^{2^x -1 }|\varphi_j\rangle\;,
\end{equation}
where $j$ enumerates the different paths.
To approximate this sum, one can drop some of the terms. In the context of multiprocess simulation,
one can simply skip some of the processes.
When simulating quantum-supremacy circuits \cite{GoogleSuprem}, the $2^x$ terms tend
to have nearly-identical norms because odd amplitudes and even amplitudes selected by projections in Equation (\ref{eq:CZ}) are equally distributed. This is the worst case (otherwise, we would have dropped
terms with lower norms), so we assume $\| |\varphi_j \rangle \| = 2^{-x/2} ~~ \forall j$. Furthermore, each path corresponds to a different quantum random circuit, and therefore their output states are almost orthogonal, $|\langle \varphi_j | \varphi_k \rangle|^2 \simeq 2^{- x \cdot n}$ for $j \ne k$.
 For requested simulation fidelity $0<f\leq1$, we include $f 2^x$ terms out of $2^x$
 \begin{equation}
 |\psi_a\rangle = \sum_{j=0}^{f 2^x -1 } |\varphi_j\rangle,
 ~~\mathrm{so}~~ \langle \psi_a |\psi_a\rangle= f 2^{-x} \textrm{(see Figure~\ref{fig:norm_fluc})}\;.
 \label{eq:fid}
 \end{equation}
Fidelity can then be estimated as
\begin{equation}
\frac{ \langle \psi |\psi_a\rangle \langle \psi_a |\psi\rangle}{\langle \psi_a |\psi_a\rangle} = \frac{f^2 2^{-x}}{f 2^{-x}} = f\;,
\end{equation}
see Figure~\ref{fig:fidelity_fluc}.
Attaining fidelity $f$ with only an $f$ fraction of work is remarkable, e.g.,
quantum simulation with fidelity 0.1 is 10$\times$ faster than exact simulation.
This scaling holds up well in multiprocess simulation, as we confirmed empirically.
The specific selection as to which
 $|\varphi_j\rangle$ terms to leave in Equation (\ref{eq:fid}) does not matter
for Google quantum supremacy circuits because the norms $\| |\varphi_j \rangle \|$ are nearly identical (we checked this). In general, dropping terms with smaller norms can yield a better-than-linear
tradeoff between fidelity and runtime.

Approximate simulation methods in \cite{bravyi_improved_2016, bravyi_2018} do not  appear particularly promising on quantum-supremacy circuits \cite{GoogleSuprem}, which have numerous T gates (see Tables \ref{tab:results-easy} and \ref{tab:results-hard})
. The results in \cite{bravyi_improved_2016, bravyi_2018} differ from ours in another essential way. Their runtime scales with $1/\epsilon^2$ for circuit error-rate $\epsilon>0$, which is beneficial for {\em small-error simulation}. Our scaling, linear in $f$, is beneficial for {\em small-fidelity simulation} and
 near-term quantum-supremacy experiments. 
 Cloud-based approximate simulation is discussed in Section~\ref{sec:perf}.

\ \\
\noindent {\bf Any cross-block multiqubit gate} can be handled with our techniques by replacing the decomposition in Equation (\ref{eq:CZ}) by an operator Schmidt decomposition of the gate, ensuring the minimal number of terms. The linear scaling of runtime with fidelity does not depend on this Schmidt rank.

\begin{table}[tb]
\begin{center}
\begin{tabular}{|c|r|r|c|r|r|r|r|r|r|}
\hline
  Qubit  & Circuit  & \multicolumn{4}{|c|}{Gate counts} & RAM & Runtime   &  Cost   \\
  array  & depth & total & T & CZ & xCZ &  GiB     & hr            & \$ \\
  \hline
  $6\times 7$ & $1+25$ & 600 & 131 & 222 & 18 & 8.58 & 4.6  &  3.34  \\
  $7\times6$ & $1+25+1$ & 711& 157 & 224 & 18 & 8.58 & 6.0  &  4.32  \\
  $9\times5$ &$1+25$ & 643 & 139 & 238 & 15 & 17.4 & 0.3 & 0.24    \\
  $9\times5$ &$1+25+1$ & 767 & 176  & 237 & 15 & 17.4 & 1.4 & 1.01    \\
 \hline
\end{tabular}
\parbox{14cm}{
\caption{\label{tab:results-easy} Exact simulation of ``easy'' quantum-supremacy circuits used in \cite{GoogleSuprem,PB} on a single Google Cloud {\tt n1-highcpu-96} server with 96 hyper-threads, equivalent to 48 cores of 2.0 GHz Intel (Skylake) CPUs. Circuit depth is given in cycles,
where the initial and final 1s represents Hadamard gates. Each simulation saved 1M amplitudes, 
requested at random (the I/O overhead is included in runtime). Cost estimates are based on \$0.72/hr pricing for preemptible VMs. Circuits with $7\times6$ and $9\times5$ qubits and depth $1+25+1$ represent refined benchmarks from Section~\ref{sec:bench}, including the circuit file {\tt inst\_7x6\_26\_0}~\cite{new_benchmarks}.  }
}
\end{center}
\vspace{-4mm}
\end{table}

\section{How many amplitudes must be calculated?}
\label{sec:sampling}

The computational task for quantum supremacy explored in this paper entails sampling from a distribution
of bistrings that is defined by the output of a given random quantum circuit~\cite{GoogleSuprem}. 
To perform this task, we now leverage our simulation techniques that provide the output probabilities $\{p(x_j)\}$ for any chosen set of bistrings $\{x_j\}$ (see Table \ref{tab:10e8}). Our {\em frugal rejection sampling} yields $\ell$ bistrings from the appropriate distribution (on average) using only $10 \ell$ output probabilities. For the fidelity considered here, a quantum computer must sample $\sim 10^6$ bitstrings to enable cross-entropy benchmarking~\cite{GoogleSuprem}, so a simulation must find $\sim 10^7$ amplitudes/probabilities. 

\noindent {\bf Basic rejection sampling.}
A quantum computer running an $n$-qubit circuit from \cite{GoogleSuprem} may yield any one of $N=2^n$ bitstrings on the output. However, outcome probabilities $\{p(x_j)\}$ for sufficiently deep circuits \cite{GoogleSuprem} follow an exponential (Porter-Thomas) form $\frac  1 N \sum_x \delta (p(x) - p ) = N e^{-N p}$~\cite{GoogleSuprem}. Figures~\ref{fig:pt} and \ref{fig:apt} show that our simulations --- both exact and approximate --- reproduce this trend. As a first approximation of quantum sampling, a competing classical computer might select a bitstring $x_j$ uniformly at random, calculate its probability $p(x_j)$ by simulation, and accept $x_j$ with probability ${p(x_j)N/M}\leq 1$, repeating the entire process until acceptance. This {\em rejection sampling} assumes a bound $M \ge \max_x p(x) N$ and requires an average of $M$ probabilities $p(x)$ per accepted bitstring. It can be implemented by Algorithm 1 in Figure \ref{fig:sampling}.
We can estimate the expected number $\epsilon$ of probabilities $p(x) > M/N$ using the Porter-Thomas distribution as follows
\vspace{-1mm}
\begin{equation}
  \mathrm{ \bf Exp} \Big[ \sum_{p(x) \ge M/N}  1 \Big] =
    \epsilon =  N \int_{M}^\infty e^{-t} d t = N e^{- M}\;.
\end{equation}
Hence for given $\epsilon$ we can choose $M = \ln(N/\epsilon)$. For $\epsilon = 10^{-3}$ and $n = 49$, we need $M = \lceil \ln(10^3 2^{49}) \rceil =  41$ probabilities per bitstring. 

\noindent {\bf Frugal rejection sampling.} To reduce the number $M$ of probabilities per output bistring to $M'$ without significantly increasing the error, we set $M'(\varepsilon)$ so that 
$\sum_{x : p(x) > M' /N}^N p(x) \le \varepsilon$. The distribution $\tilde{p}(x)$ from which Algorithm 1  samples has statistical variational distance from the true distribution~\cite{neville_classical_2017} 
\vspace{-1mm}
\begin{equation}
  \frac 1 2 \sum_x | p(x) - \tilde{p}  (x)| = 
     \mathrm{ \bf Exp} \Big[\sum_{x : p(x) > M' /N} p(x)\Big] ~\le ~ \varepsilon\;.
\end{equation}
To get a tight bound on the number of probabilities needed per bitstring,
we use Porter-Thomas statistics~\footnote{In comparison, boson sampling requires
$\sim 100$ probabilities per bistring~\cite{neville_classical_2017}, despite the
use of Metropolised independence sampling.}
\begin{equation}
    \mathrm{ \bf Exp}  \Big[
\sum_{x : p(x) > M' /N} p(x)  \Big] = \int_{M'}^\infty t e^{-t} d t = e^{- M'} (1+M')\;.
\end{equation}
The statistical distance decreases exponentially with the number of probabilities $M'$ per bitstring sample. $M' = 10$ gives statistical distance equal to $\varepsilon = 5 \cdot 10^{-4}$, independent of the number 
of qubits. 
To further decrease $\varepsilon$, we skip Step 4 in Algorithm 1 in Figure \ref{fig:sampling} 
and always accept $x_j$ when $p(x_j) \ge M' / N$ in Step 4 in Algorithm 2. In other words,
instead of ignoring high-probability bitstrings, we sample them with a lower probability. 
Extending earlier calculations gives
\vspace{-1mm}
\begin{equation}
  \frac 1 2 \sum_x | p(x) - \tilde{p}  (x)| = 2 \exp\left( - { M' \over 1 - e^{- M'}}\right)\;.
 \vspace{-1mm}
\end{equation}
For $M' = 10$, this reduces the statistical distance (between $p(x)$ and $\tilde{p}(x)$)
to $9 \cdot 10^{-5}$. 

Our {\em frugal rejection-sampling} can be applied without assuming Porter-Thomas statistics.
For such uses, one cannot rely on analytical error bounds, but can calculate the sampling error
(for a given $M'$) for each batch of $\ell M'$ probability values. 
We numerically obtain
$\varepsilon = \sum_{x : p(x) > 10 /N} p(x) = (4.4 \pm 0.7) \cdot 10^{-4}$ in the case of Figure~\ref{fig:pt} (exact simulation)~\footnote{The errors were estimated by bootstrapping.}, and $\varepsilon = (5.3 \pm 0.7) \cdot 10^{-4}$ in the case of
Figure~\ref{fig:apt} (approximate simulation).
Since each figure shows a good fit to Porther-Thomas, we can compare these error estimates
to the analytical prediction $5 \cdot 10^{-4}$ based on the Porter-Thomas distribution.
These errors and the runtime of rejection sampling are negligible in the context of
approximate circuit simulation, where $\ell M'$ probabilities are produced per run for Algorithm 2 
from Figure 1. Then $\ell \pm \sqrt{\ell M}$ bitstrings are generated, as seen from McDiarmid's 
inequality.

\begin{figure}[tb]
\begin{center}
\begin{tabular}{|c|c|}
\hline
\sc Algorithm 1 (Basic) & \sc Algorithm 2 (Frugal) \\
\hline
\parbox{7.3cm}{
\begin{enumerate}
\item $M = \lceil \ln (N / \epsilon) \rceil$
\item Sample a subset of distinct $\ell M$ bitstrings $\{x_j\}$ uniformly at random.
\item Calculate $\ell M$ probabilities $\{p(x_j)\}$ at once.
\item Remove bitstrings $x_j$ with ${p(x_j) > M/N}$.
\item For each remaining $j$: accept $x_j$ with probability ${p(x_j) N/M}$.
\end{enumerate}
}
&
\parbox{7.4cm}{
\begin{enumerate}
\item Pick $ M'$ so that $\sum_{x : p(x) > M' /N}^N p(x) \le \varepsilon$.
\item Sample a subset of distinct $\ell M'$ bitstrings $\{x_j\}$ uniformly at random.
\item Calculate $\ell M'$ probabilities $\{p(x_j)\}$ at once.
\item For each $j$: accept $x_j$ with probability $\min\{1, p(x_j) N/M'\}$.
\end{enumerate}
\ \\
}
\\
\hline
\end{tabular}
\parbox{14cm}{\caption{\label{fig:sampling} Sampling $\ell$ bitstrings from a subset of size $\ell M$ (or $\ell M'$)
with known probabilities. The  two algorithms differ in how they $(i)$ determine set size and $(ii)$ handle high probabilities.
}}
\end{center}
\vspace{-6mm}
\end{figure}

\section{Simulation results}
\label{sec:perf}

The infrastructure used by our simulation is described in 
Sec.~\ref{sec:methods}. 
To put our results in perspective, Appendices~\ref{app:QDK} and \ref{app:QASM} compare Rollright (without approximation) to the most recent simulators from Microsoft and IBM on circuits up to 30 qubits. Rollright shows speed advantage of $31\times$ and $16\times$ over Microsoft QDK and IBM QISKit-Terra/QASM, respectively. Running QISKit-Terra on Power9 servers with NVIDIA GPUs through the IBM Q service decreases Rollright's speed advantage to $12.8\times$, but for a 32-qubit circuit Rollright is $877\times$ faster. Without using GPUs, Rollright simulates 30-qubit circuits on a MacBook Pro at least twice 
as fast as QISKit on GPUs. Rollright's memory advantage over Microsoft and IBM is $886\times$ and $590\times$. Appendix~\ref{app:64_22} shows that Rollright considerably outperforms
the simulator from \cite{64q} on 64-qubit circuits of depth 22.

For simulations of ``easy'' quantum circuits (42 and 45 qubits), we use 4 threads per process with 22 parallel processes on one {\tt n1-highcpu-96} server with 96 vCPUs.~\footnote{On Google Cloud, a virtual CPU is implemented as a single hardware hyper-thread.} 
Simulations for up to 45 qubits were performed in exact mode,
 in configurations that match those previously reported, so as to facilitate comparisons (the gate counts
 do not always match exactly, but these differences have only minor impact on results). As implied by
 Table \ref{tab:results-easy}, we use much more modest resources than supercomputer-based simulations from 2016 and 2017, and achieve lower runtimes than cluster-based simulations. In particular, a 45-qubit simulation of depth $1+25$, previously performed on a supercomputer with 0.5PB RAM \cite{PB}, took \$0.24 and 17.4 GiB RAM with our algorithms. This illustrates rapid progress in the understanding and availability of quantum circuit simulation. 

\begin{figure*}[t!]
    \centering
    \begin{subfigure}[b]{0.45\textwidth}
        \centering
        \includegraphics[width=\textwidth]{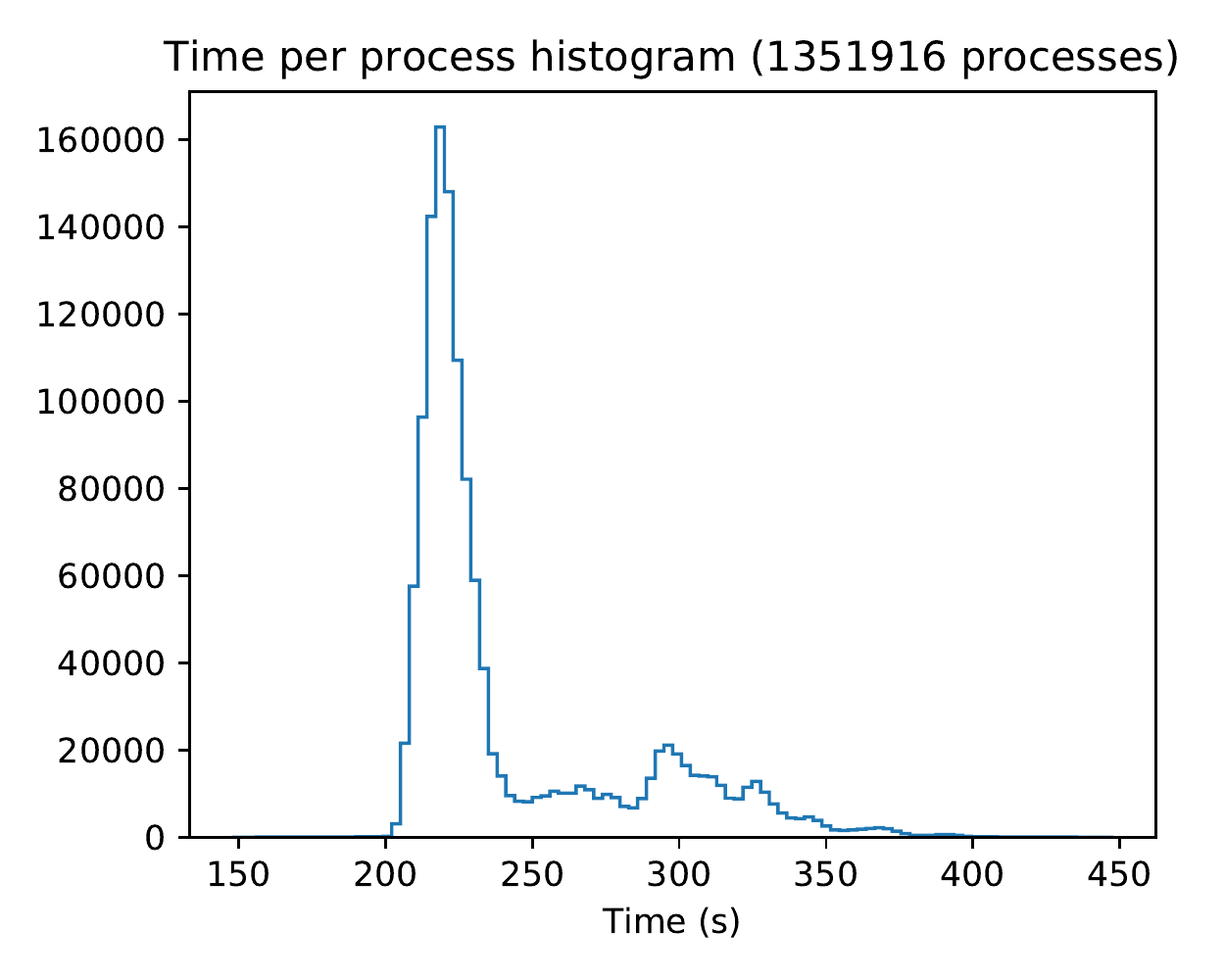}
        \vspace{-8mm}
        \caption{\label{fig:times_7x7} {\tt inst\_7x7\_41\_0} }
    \end{subfigure}
    ~
    \begin{subfigure}[b]{0.45\textwidth}
        \centering
        \includegraphics[width=\textwidth]{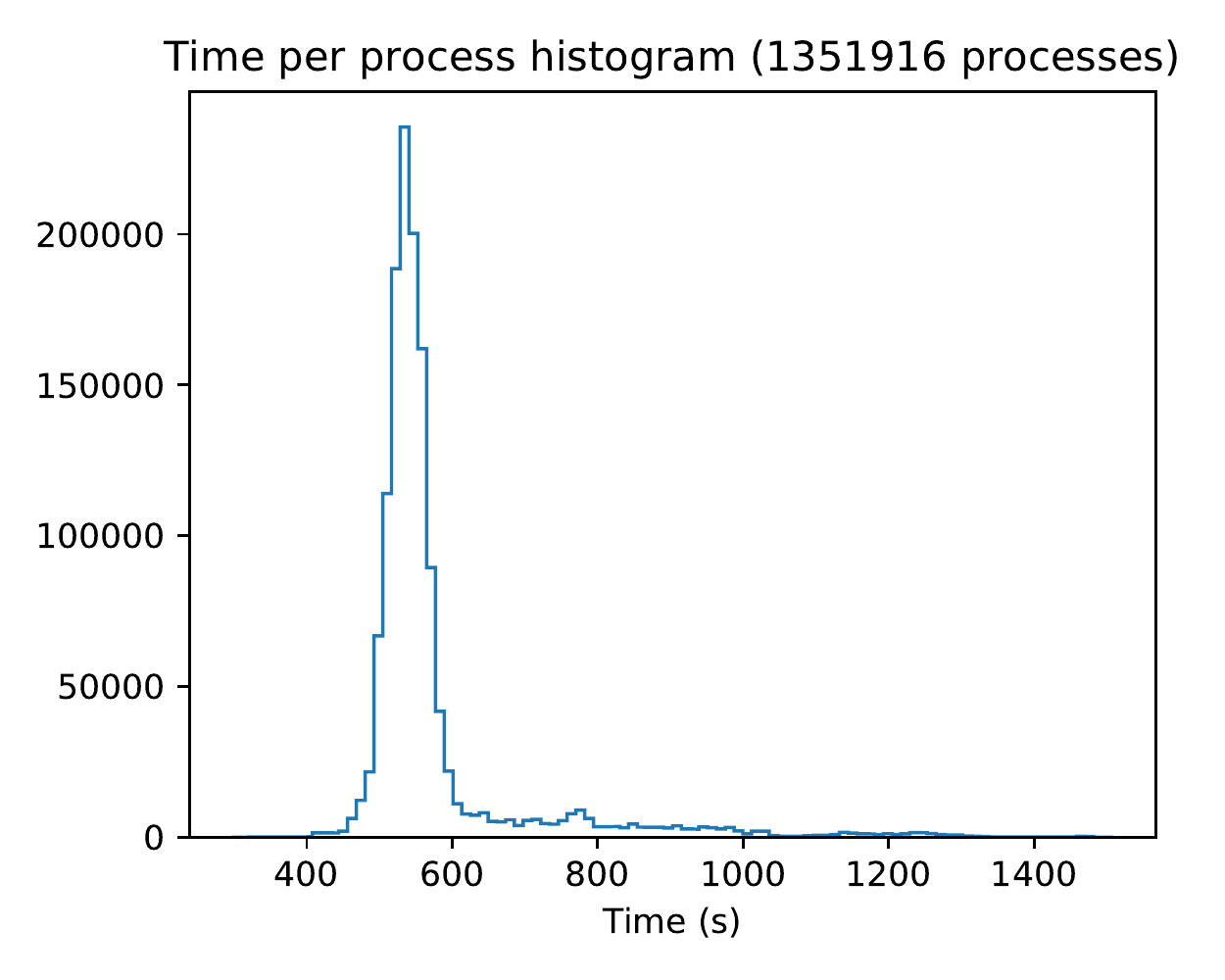}
        \vspace{-8mm}
        \caption{\label{fig:times_7x8} {\tt inst\_7x8\_41\_0}}
    \end{subfigure}
    \caption{\label{fig:49and56} Single-process runtimes when simulating ($a$) a $7\times7$-qubit circuit with depth 1+40+1, and ($b$) a $7\times8$-qubit circuit with depth $1+40+1$. Hardware-performance variations are possibly due to CPU diversity and transient loads at cloud compute nodes. Process runtimes on the same node tend to be very similar. Each process computes $2^7$ branches, corresponding to $x_b = 7$ in Eq.~(\ref{eq:t_tot}), for $10^6$ amplitudes using 8 threads.}
    \vspace{-4mm}
  \end{figure*}

\ \\
\noindent
{\bf Runtime variation} is important to track for two reasons.
\begin{itemize}
\item Simulators in some prior work~\cite{chen_classical_2018} appear exponentially sensitive to quantum-supremacy circuit instances, possibly because they exploit easy-to-simulate gate sequences that can be removed (see Section~\ref{sec:bench}). This is not the case in our work.
For different circuits with the same depth and numbers of qubits, same placement of xCZs and similar number of gates, the runtime of our simulator is also similar. For a given circuit,
the amount of work performed by individual branches of simulation and their runtimes on a given CPU do not exceed mean values for that CPU by more than 10-20\%.
\item Using
 servers, especially in a shared Cloud Computing environment, exposes simulations to hardware-performance variations possibly due to CPU diversity and transient loads at compute nodes (jobs by other users can schedule on the same compute nodes), see Figure~\ref{fig:49and56}.
\end{itemize}

\noindent {\bf Approximate simulation of ``hard'' quantum circuits.}
Simulations for $7\times7$ and $7\times8$ qubit arrays are more difficult than simulations we used
 for benchmarking purposes above, and are performed to a greater circuit depth in our work,  $1+39+1$ and $1+40+1$, including
 the initial and the final layers of Hadamard gates~\cite{new_benchmarks}. Depth of at least 40 (plus Hadamards) was suggested \cite{GoogleSuprem,GoogleBucket} as sufficient to be hard for simulations optimized for low depth circuits. It is easier to simulate circuits on oblong arrays than on square arrays, although specific algorithms may be affected in different ways \cite{GoogleBucket}.
 For our simulation,
 \begin{itemize}
\item  we choose a cut with 7 xCZ gates every 8 cycles,
\item  the $7\times7$ array does not admit a balanced cut, and
 the blocks formed by the most balanced, smallest cut contain 28 and 21 qubits, whereas the $7\times8$ array admits an even 28 + 28 partition.
 \end{itemize}
 This level of simulation difficulty offers a good opportunity to evaluate our approximate simulation technique.
 Therefore, we use $f=10^{-2}$ for the $7\times7$ array depth  $1+39+1$ , and $f=1/196$ for depth $1+40+1$ .
Distributed simulations were performed on shared physical hardware in Google Cloud,
using up to 625 {\tt n1-highcpu-32} virtual servers for $7 \times 7$ qubits and 625 {\tt n1-highmem-32} servers for $7\times8$ qubits, with 32 vCPUs each. Key results are reported in Table \ref{tab:results-hard},
and details are available in appendices. These results can be used to estimate resources for different fidelity values $f$ because runtime scales linearly with $f$. The linear scaling does not depend on the number of qubits or circuit depth.

We also performed a simulation of a circuit with $7\times 7$ qubits and depth $1+48+1$ with fidelity $2^{-22}$ using 512 {\tt n1-highcpu-32} virtual servers with an estimated cost \$52. Using the linear scaling of computational cost with fidelity, this implies a cost of one million dollars for 0.5\% fidelity.

\begin{table}[b]
\begin{center}
\begin{tabular}{|c|r|r|c|r|r|r|r|r|r|r|}
\hline
  Qubit &  Circuit  & \multicolumn{4}{|c}{Gate counts} & Fide- & RAM & \multicolumn{2}{|c|}{Runtime, hr} &  Cost   \\
  array & depth & total & T & CZ & xCZ &  lity    &    TiB  &  clock  &  billable & \$ \\
  \hline
    $7\times7$ & $1+39+1$ & 1252 & 286 & 410 &  31 & $ 1\%$& 15 & 35.2  & 2.2e+04   & 5218  \\
  $7\times7$ &  $1+40+1$ & 1280 & 293 & 420 & 35 & $ 0.51\%$& 15   & 58.2  &  3.6e+04   & 8734  \\
  $7\times8$ &  $1+40+1$ & 1466 & 331 & 485 & 35  & $ 0.51\%$& 30 & 140.7   & 8.8e+04 & 35184  \\
 \hline
\end{tabular}
\parbox{14cm}{
\caption{\label{tab:results-hard} Approximate simulation of ``hard'' quantum-supremacy circuits \cite{new_benchmarks} on Google Cloud servers. Here we use v2 benchmarks from Section~\ref{sec:bench} that address weaknesses found in v1 benchmarks~\cite{GoogleSuprem}.  The circuits are publicly available at~\cite{new_benchmarks}. Circuit depth is given in cycles, including the initial and final Hadamard gates on each qubit. Each simulation saved 1M amplitudes requested at random. The first simulation used circuit file {\tt inst\_7x7\_40\_0} and 617 {\tt n1-highcpu-32} machines, with cost estimate based on $\$0.24$/hr pricing for preemptible VMs. The second simulation used circuit file {\tt inst\_7x7\_41\_0} and 625 {\tt n1-highcpu-32} machines. The last simulation used circuit file {\tt inst\_7x8\_41\_0} and 625 {\tt n1-highmem-32} machines with cost estimates based on $\$0.40$/hr pricing for preemptible VMs.}
}
\end{center}
\vspace{-4mm}
\end{table}

\begin{table}[tb]
\begin{center}
\begin{tabular}{l|rrrrr}
Amplitudes saved & 1       & $10^3$ & $10^6$ & $10^7$ & $10^8$ \\ 
\hline
Mean time, $s$     & 185.7  & 184.0    & 217.7    & 512.6    & 3952 \\
Std. deviation     & 6.0     &  5.9       & 7.7       &  6.0       & 58.4  \\ 
\hline
\end{tabular}

\parbox{14cm}{\caption{\label{tab:10e8} The scaling of simulation runtimes when saving $1-10^8$
amplitudes for a 49-qubit circuit of depth 1+40+1. Here we used 10 virtual machines,
each with four concurrent processes running eight threads each (a full simulation scales the same way).
}}
\end{center}
\vspace{-4mm}
\end{table}

\ \\
\noindent {\bf Scaling simulation to produce multiple bitstring samples.} As discussed in Section \ref{sec:sampling}, we strive to emulate a quantum computer that samples bitstrings from a circuit-defined
distribution. The fidelity $f$ of the output of a noisy quantum computer can be estimated with cross-entropy benchmarking~\cite{GoogleSuprem}, which requires sampling $\sim 1/f^2$ bitstrings. For the fidelities considered in this paper, consistent with the noise estimates for near-term quantum computers, this requires to produce at least one million bitstrings or repeated runs.

To produce bitstring samples, we first select a number of bitstrings uniformly at random, then simulate the circuit to save the bitstring probabilities, and then use methods from Section~\ref{sec:sampling}
to produce a bitstring per ten probabilities on average. Therefore, to match  a million runs of a quantum computer, a simulator would produce ten million amplitudes with randomly selected indices. For algorithms that produce one amplitude at a time \cite{GoogleBucket,chen_classical_2018}, this implies a 10$^7\times$ slow-down to produce $10^6$ bitstrings. 
In contrast, our simulator produces multiple amplitudes in a batch. Table \ref{tab:10e8} shows
the scaling of runtimes when the number of amplitudes saved (with random indices) increases
from 1 to $10^8$. Saving $10^6$ amplitudes leads to a 17\% slowdown because the bottleneck
is in the simulation itself. Saving $10^7$ amplitudes leads to a $2.76\times$ slowdown. Producing amplitudes in a batch gives us a $10^7/2.76 = 3.6\times10^6$ speedup over producing one amplitude at a time.

\section{Refined quantum-supremacy benchmarks}
\label{sec:bench}

We now review the difficulty of Quantum Supremacy benchmarks for sampling the output distribution of random quantum circuits. Analysis in \cite{GoogleSuprem} suggests circuits with at least $7 \times 7$ qubits, as the quantum state of such a system is too large for typical Schr\"odinger-style simulations. With the same $7 \times 7$ array,
circuit depth must be at least $1+40+1$, given the exponential growth of simulation difficulty with depth for  simulation by variable-elimination~\cite{GoogleSuprem,GoogleBucket}. Note that the simulation of sampling requires calculating a large number of amplitudes requested at random~\cite{neville_classical_2017}, whereas variable elimination tends to produce one amplitude at a time. We have shown in previous sections how to simulate $7 \times 7$ circuits with CZ gates of depth $1+40+1$ trading computational cost for fidelity. This suggests revising the initially proposed benchmarks \cite{GoogleSuprem}.

 Quantum Supremacy entails an inherently adversarial protocol that asymmetrically favors quantum computers --- a computational problem is being selected that can be solved by quantum yet not classical computers \cite{aaronson2011computational,bremner2015average,Preskill,bremner16,GoogleSuprem,NatureSupremacy,aaronson2016complexity,neill_blueprint_2017,bouland_quantum_2018}. Reliably defeating Quantum Supremacy requires more than a handful of opportunistic simulations as one has to anticipate modifications of problem instances that complicate simulation more than they complicate quantum evolution. It is sometimes easy to confuse Quantum Supremacy with almost the opposite, i.e., showing that classical computers can solve tasks that present-day quantum computers cannot. The later is trivial and can be demonstrated on many common tasks for which classical software and hardware excel, while quantum computers have no algorithmic advantage and are currently much smaller and error-prone. For example, sorting $2^n$ numbers requires $\Omega(n2^n)$ time on both classical and quantum computers~\cite{hoyer2002quantum} and can be accomplished just as quickly. Of course, such statements do not preclude Quantum Supremacy. On the other hand, being able to quickly simulate only some circuits of a kind~\cite{gottesman1998heisenberg,bravyi_improved_2016, bravyi_2018} would not negate Quantum Supremacy, e.g., the work in~\cite{chen_classical_2018} attempts to simulate a number of similar circuits but then abandons the ones that time out. The authors of~\cite{64q} change the layout of CZ gates to make the simulation approximately 16 times faster at depth $1+22$. 
Simulations in \cite{Taihu,chen_classical_2018} attain additional depth by omitting the final layer of Hadamard gates, which makes the CZs in preceding cycles unnecessary. Moreover, the work in~\cite{chen_classical_2018} calculates a single amplitude at a time, which is insufficient for sampling.
Finding unexpected simulation shortcuts in existing quantum-supremacy benchmarks \cite{GoogleSuprem} is also of limited value if these shortcuts can be invalidated.  To this end, the work in \cite{Taihu} describes how to simulate eight additional cycles in some cases by exploiting sequences of diagonal gates \verb+CZ - T - CZ+.  Avoiding such ``easy'' sequences during benchmark generation does not affect the execution on a quantum computer. Thus, Google benchmarks have been revised to
 \begin{itemize}
\item replace every T gate appearing after a CZ gate with a non-diagonal gate,
\item explicitly include a cycle of Hadamard gates before measurement,
\item reorder cycles so that ``horizontal'' two-qubit gates alternate with ``vertical'' two-qubit gates.
\end{itemize}
While keeping the same gate family, these minimal changes complicate many approaches to simulation. 
As seen in Tables \ref{tab:results-easy} and \ref{tab:results-hard}, the benchmarks retain numerous non-Clifford gates to hamper stabilizer-based simulation \cite{bravyi_improved_2016, bravyi_2018}.
Revised circuit benchmarks are publicly available~\cite{new_benchmarks} to facilitate fair runtime comparisons among different simulation methods. To ensure the replicability of results, we recommend mentioning specific circuit files, as we do in our work. The performance of our simulator on revised benchmarks
is compared to its performance on older benchmarks (see Table \ref{tab:results-easy}). The slowdown is moderate and primarily due to the slight increase in the number of gates, especially the Hadamards at
the end of the circuit.

\section{Harder benchmarks through two-qubit gates}
\label{sec:2q}

In considering more drastic changes to circuit benchmarks, we estimate the impact of upgrading the gate library on simulation difficulty and on specific algorithms used in our work.
 To this end, replacing existing quantum gates within tensor blocks might require additional provisions for simulators that implement speedups for particular gate types, but would result only in a constant-times runtime difference. However, replacing cross-block CZ (xCZ) gates with more complicated gates can increase the branching factor in our multiprocessor simulation.  The 4x4 matrix of a generic two-qubit gate can be decomposed into a sum of four tensor products of one-qubit gates
\begin{equation}
\label{eq:4way}
\left(
   \begin{array}{cc}
     A_{00} & A_{01} \\
     A_{10} & A_{11} \\
   \end{array}
 \right)
  = \rP_{0}\otimes A_{00} +  \rP_{1}\otimes A_{11} +  |0\rangle \langle 1| \otimes A_{01} +|1\rangle\langle 0|\otimes A_{10}
\end{equation}
where the $A$ terms represent 2x2 blocks of A.\footnote{A similar decomposition can be written out with the projections appearing in the second tensor component.}
 For example,
\begin{equation}
\label{eq:iSWAP1}
  {\rm iSWAP} = \rP_{0}\otimes \rP_{0} + \rP_{1}\otimes \rP_{1}+  i |0\rangle \langle 1|\otimes |1\rangle\langle 0| + i |1\rangle\langle 0|\otimes |0\rangle \langle 1|\;.
\end{equation}
Note that non-unitary gates pose no difficulties to common simulation algorithms, and projections
are particularly convenient because they create numerous intermediate zero amplitudes that can be skipped
by simulators that track patterns of nonzeros (cf. Equation (\ref{eq:iSWAP2})). Lemma 1 in \cite{nielsen2003quantum} shows that such decompositions cannot be shorter than the operator Schmidt decomposition (which must have rank 1, 2 or 4 for a unitary operator). The operator Schmidt decomposition for iSWAP
\begin{equation}
\label{eq:iSWAP2}
  2 \cdot {\rm iSWAP} = \rI \otimes \rI + i \rX\otimes \rX +
 + i \rY\otimes \rY + \rZ\otimes \rZ
\end{equation}
proves that the decomposition in Equation (\ref{eq:iSWAP1}) is minimal.
If all CZ gates are replaced with iSWAPs, the branching factor in multiprocess simulation increases from two to four, halving the circuit depth that can be handled within a given amount of time. Similar, if not greater, complications hamper algorithms which map the circuit to undirected graphical models and use variable elimination~\cite{GoogleBucket,chen_classical_2018}. 

Using an arbitrary two-qubit gate can slow down simulation of individual qubit blocks and also decrease the number of intermediate zero amplitudes, but the iSWAP gate has been well-studied and implemented in leading quantum chips \cite{neeley2010generation,mckay2016universal}. Moreover, an arbitrary two-qubit gate may be easier to approximate than an iSWAP by truncating the least-norm terms in its operator Schmidt decomposition. To create a new benchmark suite, while preserving the structure of existing benchmarks, we replace every CZ gate with an iSWAP gate. This benchmark suite is also available at~\cite{new_benchmarks}.

In addition to considering more general one-qubit and two-qubit gates, the benchmarks can be modified
to include two-qubit gates that couple arbitrary (non-adjacent) qubits. When such gates are applied within tensor partitions, our simulator will treat them just as efficiently as it treats gates in existing benchmarks. However, when such gates cross partitions, they will contribute to the branching factor of simulation. Nevertheless, our work focuses on simulating realistic quantum-circuit architectures, such as those pursued by IBM, Google, Rigetti and others. To this end, existing benchmarks are sufficient.

\section{Conclusions}
Universal quantum computers based on several different technologies started scaling beyond a handful of qubits around 2011. Today, detailed reports are published of programmable quantum-circuit computers with 22 qubits, while 50-qubit systems and larger are under testing~\cite{ibm_2017,intel_18,google_2018}. Two-qubit gate fidelities have improved within the 0.9 - 0.995 range \cite{barends_superconducting_2014,barends_digitized_2015,kelly_state_2015}, but continue to limit implementable circuit depth. Numerical simulations traditionally focused on straightforward full-wave-function kernels, where progress from 42 qubits in 2010 to 45 qubits in 2018~\cite{de_raedt_massively_2018} had been bounded by exponential memory scaling. The more recent focus on limited-depth quantum circuits extended supercomputer-based simulations well past 50 qubits \cite{IBM,GoogleBucket,64q,Taihu}, facilitated the use of GPUs and distributed cluster without interprocess communication \cite{64q}, and enabled meaningful simulations on single servers \cite{GoogleBucket, 64q} for up to 100 qubits.

We show that full-circuit fidelity in quantum-circuit simulation can be traded off for computational resources, and a simulation can match requested circuit fidelity to compete with a given quantum computer. We also show that ($i$) competitive simulations of quantum-supremacy circuits do not require specialized systems, and ($ii$) monetary cost can be estimated for cloud-based simulations. Within a given physical-time span, supercomputers might be able to simulate larger quantum circuits than those we simulated, but perhaps at a much greater cost. 

The Schr\"odinger-Feynman simulation that we developed compares favorably to competitors:
\begin{itemize}
\item Stabilizer-based methods \cite{bravyi_improved_2016, bravyi_2018}
         are exponentially sensitive to the number of non-Clifford gates, whereas diagonal gates are easy
         for our approach. To this end, simulations in \cite{bravyi_2018} involve circuits with up to 60 such
         gates, whereas the circuits we simulate include hundreds of them. 
\item Bucket-elimination methods  \cite{GoogleBucket,chen_classical_2018} handle diagonal gates well,
        but so far produce one amplitude at a time, whereas we can produce millions. Additionally,
        bucket-elimination methods have not yet been evaluated on revised quantum-supremacy
        benchmarks from Section \ref{sec:bench}, which seem harder for this approach.
\end{itemize}
Stabilizer-based techniques have the advantage of better handling non-nearest-neighbor circuits, whereas
this generalization would impact bucket elimination methods more than it would impact our approach. However, leading quantum computers are consistent with the nearest-neighbor architecture.

Our work emphasizes rapid progress in simulation algorithms
and its implications for quantum supremacy experiments, as follows.
\begin{itemize}
\item Limited-depth quantum circuits now admit surprisingly efficient simulation on cloud computers.
\item Deep quantum circuits without error correction appear out of reach for both simulation
(due to exponentially-growing resource requirements) and quantum chips (due to the exponential accumulation of gate errors). Quantum error correction is currently out of reach for near-term quantum computers \cite{Preskill}.
\item Schr\"odinger-Feynman hybrid simulation used in our work favors conventional computers over GPUs
and does not require fast interconnect found in supercomputers.
It can be cast into an Internet-scale \verb|SETI@Home|-style effort, where transient participants all over the world contribute CPU time on their computers to a joint effort.
\item Approximate simulation of quantum circuits is surprisingly effective in practice~\footnote{Complexity-theoretic arguments against the existence of classical polynomial-time approximation algorithms for this sampling problem~\cite{GoogleSuprem} do not conflict with the algorithm in our work, as our algorithm requires exponential time.}, and this raises the bar for quantum supremacy experiments because low quantum circuit fidelity (due to gate errors) facilitates efficiency gains in approximate simulation. To this end, quantum computers incur stochastic errors in qubit initialization, measurement, one- and two-qubit gates, and gate errors tend to increase with the number of qubits. In contrast, our simulation algorithms incur deterministic inaccuracies in only a very small fraction of two-qubit gates, and the complexity of simulation scales linearly with circuit fidelity $f$.
\item We show how to interactively validate the results 
        simulations with lower-fidelity simulations.
\item Second-generation benchmarks, with iSWAP gates instead of CZ gates, appear
  significantly harder to simulate, yet implementable on a quantum computer. We estimate that changing CZ gates to iSWAP gates reduces circuit depth attainable by leading simulation methods approximately two-fold. In this context, two-qubit gate fidelity remains a key parameter to demonstrate during many-qubit operation.
\end{itemize}
Pricing quantum-supremacy simulations on cloud-computing services in terms of {\em cost per amplitude} should facilitate fair comparisons between different hardware types and can reflect the lower cost of idling general-purpose resources. Comparisons with quantum computers by cost are also possible and can help ($i$) making a long-term case for commercial quantum computers, ($ii$) pricing quantum computing resources. Such comparisons must account not only for the resources used by each simulation and each quantum computer run, but also the number of runs required to produce trustworthy results.
When sampling outputs on a quantum computer based on superconducting qubits, reported times are around 45 $ns$ per cycle with CZ gates, 25 $ns$ per cycle with only single-qubit gates~\cite{barends_superconducting_2014,kelly_state_2015}. Readout can be as fast as 140 $ns$~\cite{jeffrey2014fast}, and qubits can be initialized to a known state in a few hundred $ns$~\cite{reed2010fast,magnard2018fast}. Thus, comparisons to simulation can be performed once we know circuit fidelity values for 49 or 56 qubits, as well as the cost of quantum-computing resources. As both quantum and classical computing technologies continue to improve and become less costly, quantum supremacy remains a moving target and will require careful comparisons.

\section{Infrastructure and Performance estimation}
\label{sec:methods}
\noindent {\bf Performance estimation} for our simulation is based on the following parameters.
The qubit array is partitioned in two blocks $q = q_1 + q_2$, attempting to minimize
the larger block and reduce the number of xCZ gates ($x$). We use a single straight line cut in the qubit array for each simulation.
As explained before, we split $x=x_p + x_b$. The corresponding split for circuit depth
is $d = d_p + d_b$. The simulation is configured to save $n_a > 0$ amplitudes with specified indices.
The total runtime of all simulation processes (however many threads each process may use)
can then be estimated as
\begin{equation}
  T_{tot} (f, q_1,q_2,d_p,d_b, x_p, x_b, n_a) = C_1 f 2^{x_p} (q_1 2^{q_1} + q_2 2^{q_2}) ( d_p + C_2 2^{x_b} d_b) + C_3 2^{x_p + x_b} n_a \label{eq:t_tot}
\end{equation}
  where $C_1, C_2, C_3 > 0$ are implementation-specific constants.

 In each of the $d = d_p + d_b$ cycles, we simulate on the order of $q$ gates, of which $q_1$
 gates act on $2^{q_1}$ amplitudes and $q_2$ gates act on $2^{q_2}$ amplitudes. After the first
 $d_p$ cycles, the simulation state is saved and reused in $2^{x_b}$ branches over $d_b$ cycles.
 At the end of each branch, $n_a$ requested amplitudes are collected (unlike the work in \cite{IBM},
 our simulator can save any subset of amplitudes specified by indices). For $n_a= 10000$, the last term
 is significant for $q\leq 36$, but can be neglected when $n_a \ll 2^{q_1} + 2^{q_2}$. Given a simulation
 cluster that can run $p$ simulation processes per node on $N$ nodes, we distinguish billable time and wallclock time

\begin{equation}
  T_{\rm bill} = \omega(p) T_{\rm tot}/p ~~~~~~~~
  T_{\rm clock} = T_{\rm bill} / N
\end{equation}
where  $\omega(p)\geq 1$ reflects the slowdown of an individual process sharing a node with other $p-1$ such processes,
e.g., due to memory contention. While $\omega(1)=1$, for the more common case $16 \leq p \leq 32$, $\omega(p)$ can be in the 1.5-2 range, depending on
how much memory bandwidth a particular system offers and how much bandwidth is required by each process.
In terms of RAM usage, a single process requires
\begin{equation}
   M_{\rm proc} =  (C_4 (2^{q_1} + 2^{q_2}) + n_a) \cdot \verb+sizeof(complex)+
\end{equation}
bytes, where $C_4>0$ is a small integer. Each compute node requires $p M_{\rm proc}$ bytes,
and the entire cluster peaks at $p N M_{\rm proc}$ bytes. Circuit depth does not affect RAM requirements.

\begin{figure*}[t!]
    \centering
    \begin{subfigure}[b]{0.45\textwidth}
        \centering
        \includegraphics[width=\textwidth]{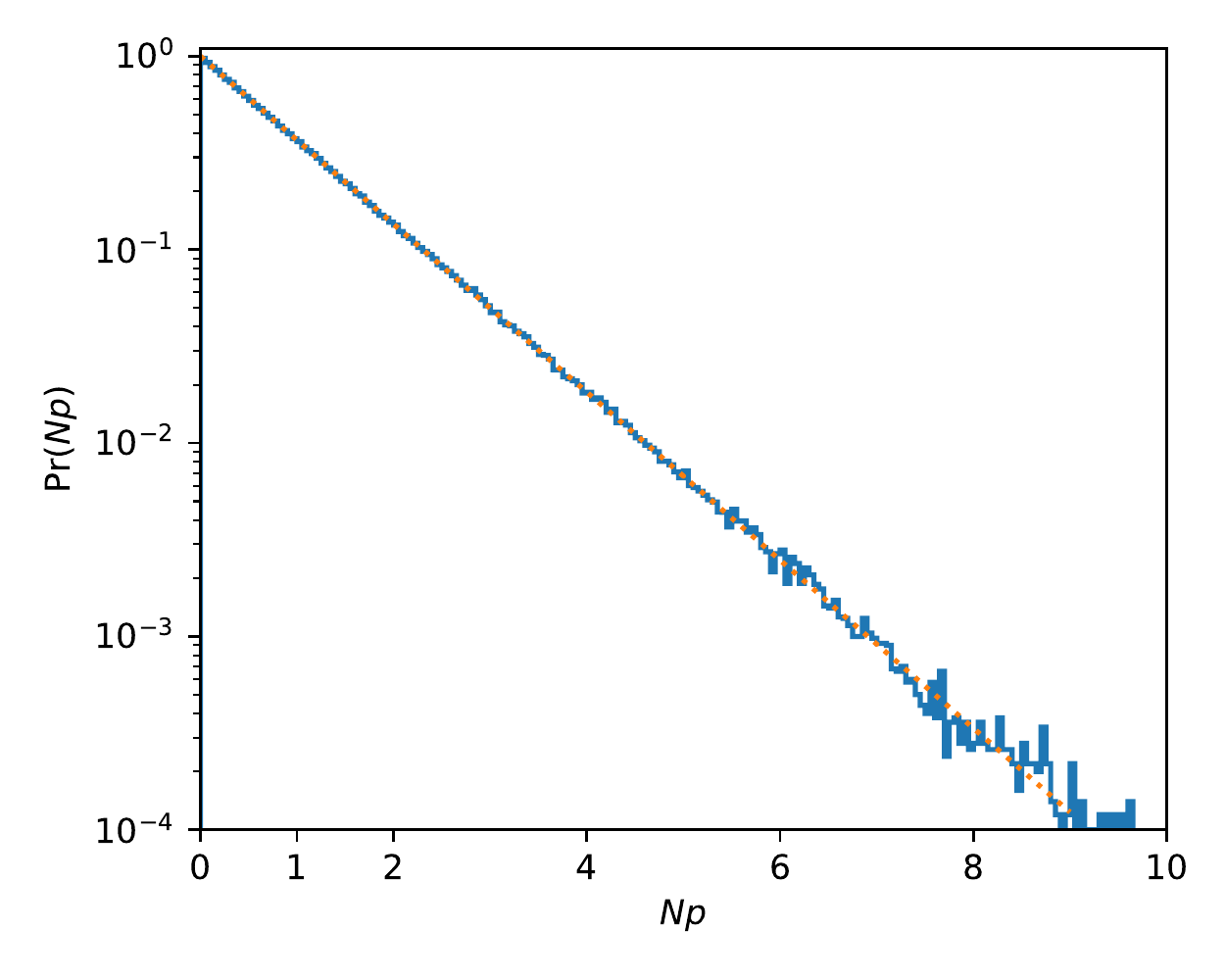}
        \caption{Match to the Porter-Thomas distribution, 45 qubits. }
        \label{fig:pt}
    \end{subfigure}
    ~
    \begin{subfigure}[b]{0.45\textwidth}
        \centering
        \includegraphics[width=\textwidth]{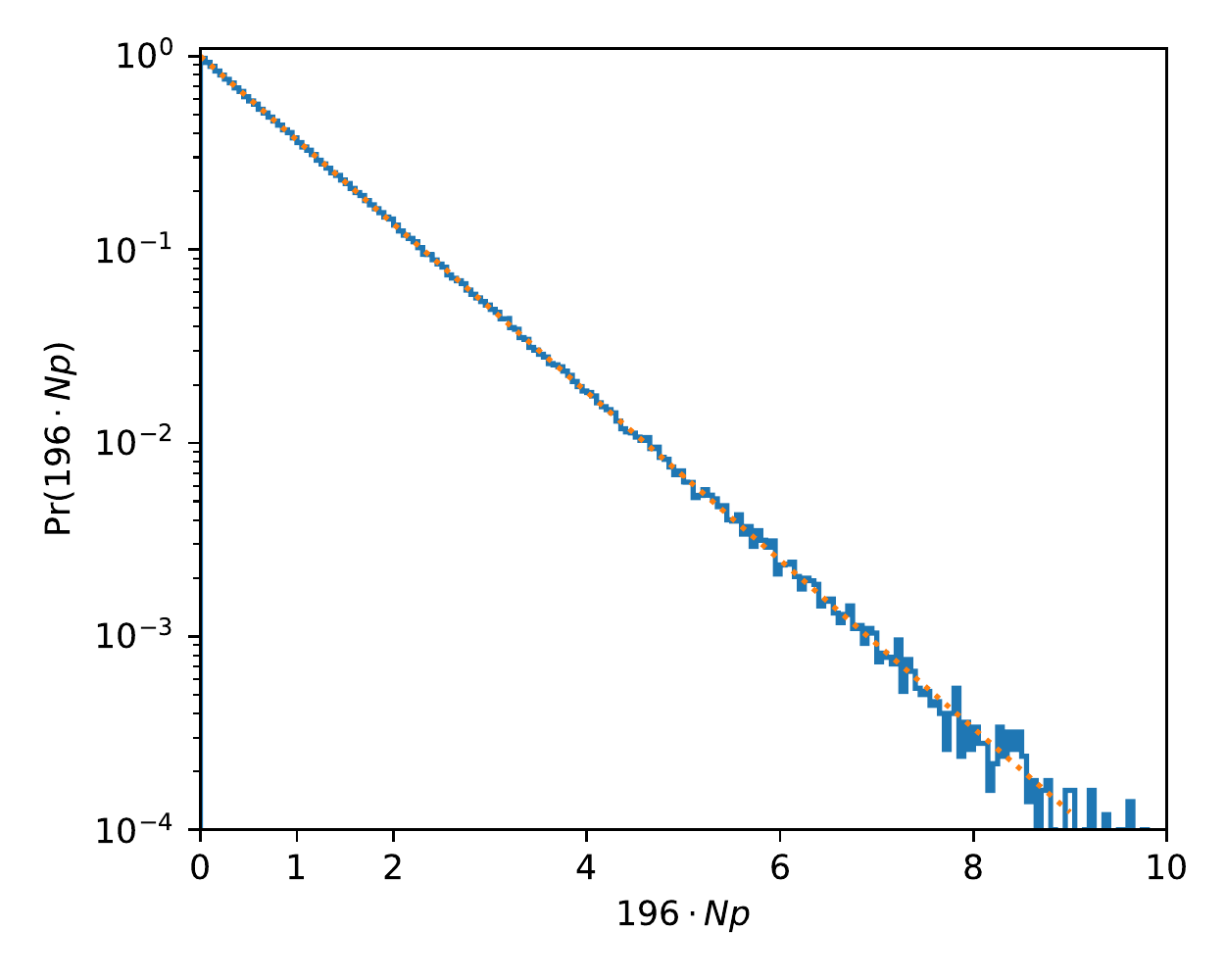}
        \caption{Porter-Thomas for 1/196 approximation, 56 qubits.}
        \label{fig:apt}
    \end{subfigure}
    \caption{Distributions of simulated bitstring probabilities for the outputs of two random circuits. (a) Exact simulation for a circuit with $9 \times 5$ qubits and depth $1+25$: bitstring probabilities match the exponential (Porter-Thomas distribution) shown with a dashed line. (b) Approximate simulation with fidelity $1/196$ for a circuit with $7 \times 8$ qubits and depth $1+40+1$, circuit file  {\tt inst\_7x8\_41\_0}~\cite{new_benchmarks}. Bitstring probabilities again match the Porter-Thomas distribution, despite rather loose approximation. }
  \end{figure*}

For the simulations reported below, the billable time and wallclock time are measured directly.

\ \\
\noindent {\bf Infrastructure used in our simulations} has been as a follows. Key algorithms are implemented  in C++11 in a package called Rollright (University of Michigan) and compiled with g++ 7.2. Our kernels use AVX-2 instructions, but not AVX-512 instructions used in \cite{PB}.  We also use python scripts for several tasks. Empirical performance is evaluated on quantum supremacy circuits \cite{GoogleSuprem,new_benchmarks}, and results are shown in Tables \ref{tab:results-easy} and \ref{tab:results-hard}, with additional details given in the appendices. Unlike prior published efforts, we use no GPUs, no supercomputers, and no high-performance node-to-node interconnect.  Software development was performed on a MacBook Pro 2017 with 16 GiB RAM, where our baseline Schr\"odinger simulation
completes a 30-qubit circuit with depth $1+26+1$ in 71 s using a little over 8 GiB of RAM. For circuits with more than 35 qubits, we use the Google Cloud, a common cloud-computing service where users rent networked servers of various types. We used several virtual server types (specified for each simulation below) with Ubuntu Linux, available to the public for fair comparisons. Most of the RAM available on those servers was not used in our simulations. For each server type, we report hourly rental cost in preemptible mode as of June 2018. Cost comparisons can be performed with other server types, adjusted for other price points, and extended to include quantum computers, so as to establish or refute quantum supremacy.

\ \\
\noindent
{\bf Practical validation of results} is a critically-important step in quantum-supremacy simulations because
subtle bugs or misguided optimizations can significantly improve runtime, while producing incorrect results.
For circuits of up to 32 qubits we validated our main Schr\"odinger simulator against several independently-developed simulators based on different algorithms, to ensure that the output amplitudes match with good accuracy (here we used the publicly available QuIDDPro \cite{Viamontes} simulator and the simulators from
IBM QISKit-Terra, as well as several in-house simulators). We then simulated circuits using our Sch\"odinger-Feynman simulator and validated results against a Schr\"odinger-only simulation, and  against a variable elimination simulation~\cite{GoogleBucket} on low-depth circuits with up to 64 qubits. Computing state norms offers a good sanity-check for both exact and approximate simulation. We also validated approximate simulation against exact simulation for low-depth circuits by producing amplitudes for the same subset of indices and estimating state-overlap (fidelity) based on these amplitudes.\footnote{We also calculated cross-entropy difference~\cite{GoogleSuprem}. This gave similar fidelity estimates, but with a greater variance of
estimation errors.} Another sanity check is to plot the distribution of probability values for a given set of
amplitudes. For exact simulation, the results should closely match the Porter-Thomas distribution, see Figure~\ref{fig:pt}. While this is not required for approximate simulation, our method also produces a Porter-Thomas distribution in this case, see Figure~\ref{fig:apt}. 

\begin{figure*}[b!]
    \centering
    \begin{subfigure}[b]{0.45\textwidth}
        \centering
        \includegraphics[width=\textwidth]{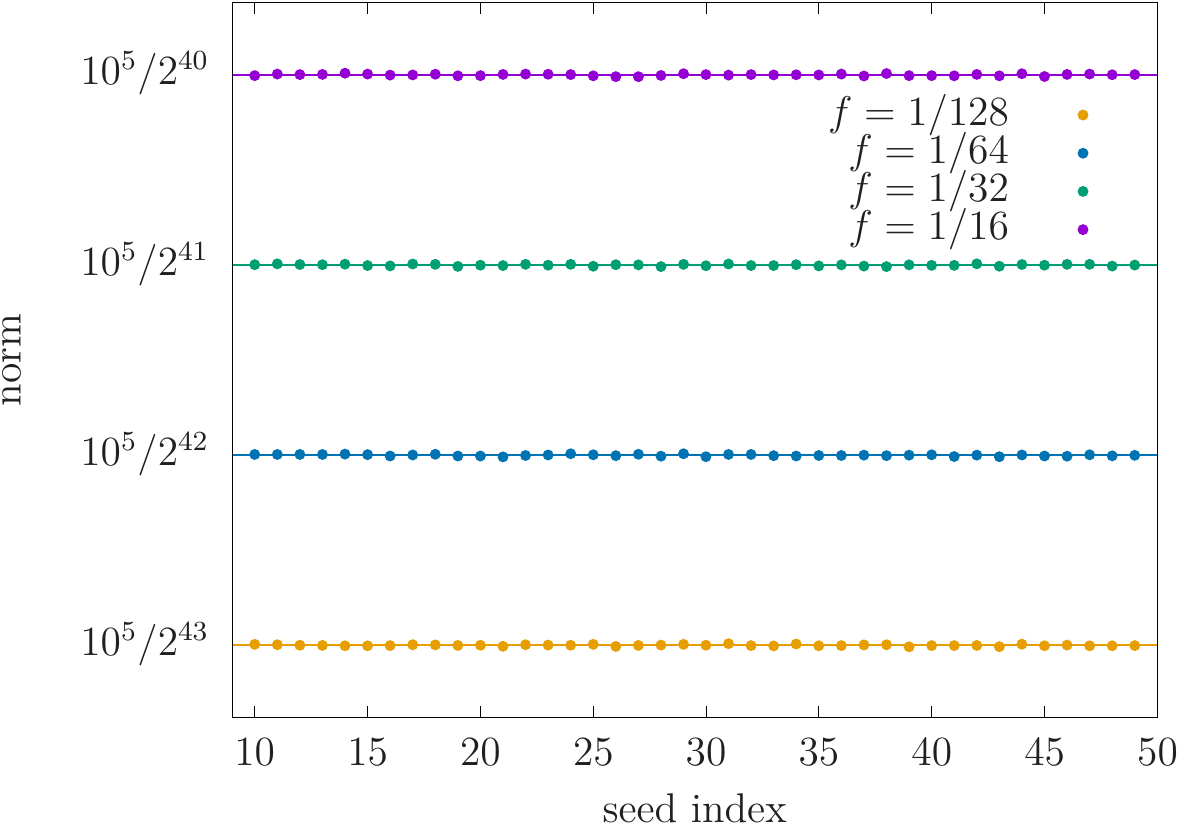}
        \caption{Norm fluctuations. }
        \label{fig:norm_fluc}
    \end{subfigure}
    ~
    \begin{subfigure}[b]{0.45\textwidth}
        \centering
        \includegraphics[width=\textwidth]{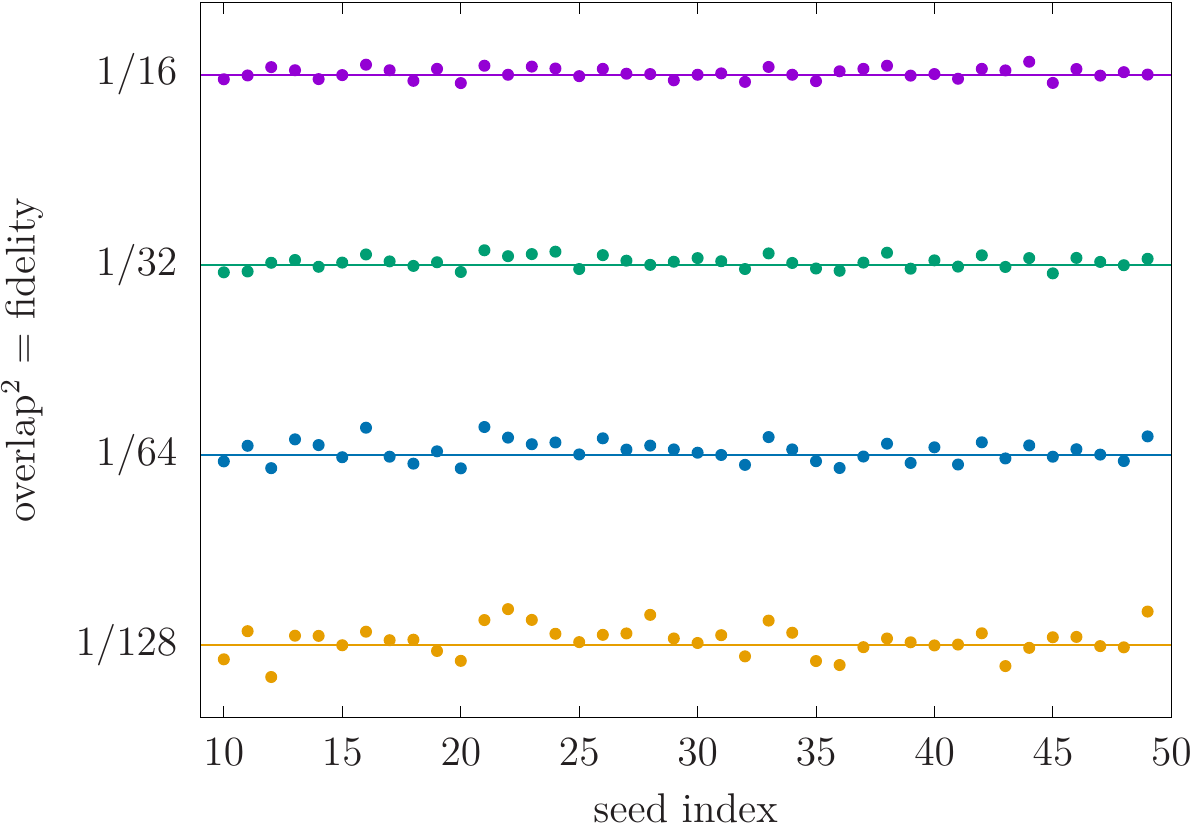}
        \caption{Fidelity fluctuations.}
        \label{fig:fidelity_fluc}
    \end{subfigure}
    \caption{Fluctuations of results when simulating the $6 \times 6$-qubit circuit {\tt inst\_6x6\_41\_0.txt} (from~\cite{new_benchmarks}) with depth $1+40+1$. In each case we choose a random fraction $f$ = fidelity of xCZ paths to obtain state $\ket{\psi_a}$ from Equation \ref{eq:fid}, then estimate its norm and fidelity using $10^5$ amplitudes. (a) The sum of  $10^5$ of the $2^{36}$ squared amplitudes of  $\ket{\psi_a}$ is $10^5/2^{36} \cdot f$, as expected. (b) The fidelity (overlap$^2$) between $\ket{\psi_a}/\|\ket{\psi_a}\|$ and the correct output is equal to the fraction $f$ of xCZs paths used.  }
  \end{figure*}

To validate the results of  approximate simulation in this work, we performed an exact simulation of  circuit {\tt inst\_6x6\_41\_0.txt} (from~\cite{new_benchmarks}) with $6 \times 6$ qubits  and depth $1+40+1$, and saved $10^5$ amplitudes. We calculated the same amplitudes using 40 different random fractions for each $f = 1/16$, 1/32, 1/64, and 1/128 of paths. Figure~\ref{fig:fidelity_fluc} shows that empirical fidelity matches the fraction of paths used. The runtimes scale linearly with $f$. We also performed an exact $7\times7$ qubit depth 17 simulation and saved 1M random amplitudes. Then we performed approximate simulation with requested fidelities $f=10^{-4}$, $f=10^{-2}$ and $f=1/64$ and saved the same amplitudes. We estimated the fidelity of approximate state by computing the inner product, and the fidelities matched the requested values to within 2\%. 
Additionally, Appendix \ref{app:validation} describes our novel technique for validating results of higher-fidelity simulation using randomized lower-fidelity simulation.

\bibliographystyle{naturemag}

\subsection*{Acknowledgments}
We acknowledge support from Kevin Kissell on running computations in Google Cloud and reviewing the manuscript,
as well as helpful comments from Edward Farhi, H\'ector Garcia, Riling Li, Dmitri Maslov, Hartmut Neven,
David Wecker, Robert Wille and Alwin Zulehner.

\appendix
\section{Comparisons to Microsoft QDK}\label{app:QDK}

To put the capabilities of our simulator in perspective, we compared it to the Microsoft Quantum Development Kit (QDK) v0.2.1806.3001 (June 30 2018).\footnote{Downloaded from {\tt https://www.microsoft.com/en-us/quantum/development-kit} in July 2018.}
This software includes a back-end circuit simulator and front-end support for the Q\# language, integrated with Microsoft Visual Studio and available on Windows, MacOS and Linux. Originally released to the public in the end of 2017, the package underwent significant improvements in February and June 2018, particularly in simulation speed. 

To compare the QDK simulator to our simulator Rollright, we used revised (v2) circuit benchmarks described in Section~\ref{sec:bench} and converted circuit descriptions into the Q\# language. In particular, circuit depth $1+26+1$ indicates layers of Hadamard gates at the beginning and at the end. Comparisons were performed on a MacBook Pro 2017 with 16 GiB RAM and Intel Core i7-7700HQ CPU (2.80GHz). Results are reported in Table \ref{tab:MSFT}. To exclude code segments from memory comparisons, we first measured max resident memory for each simulator on the 16-qubit benchmark and then used those measurements as baselines. For 24- and 25-qubit benchmarks we report the difference between max resident memory and the respective baseline. Given that the baseline includes memory used by 16-qubit data structures, these estimates are a little lower than the actual data structures, but the inaccuracy is close to 1 MiB and removed by rounding, as can be seen from Rollright data --- 128 MiB and 256 MiB are the actual sizes of amplitude vectors in Rollright. We note that Q\# uses double-precision floating-point numbers to represent amplitudes, whereas Rollright uses single-precision numbers (while keeping track of accuracy through higher-accuracy norm computations).
Hence, the memory-usage ratio should be at least 2.0 on any platform. The runtime ratios are more remarkable and are explained by two factors: ($i$) a certain slowdown due to the use of the Microsoft .NET framework, ($ii$) algorithmic differences. In particular, algorithmic differences explain why the runtime ratio grows with the number of qubits.

We note that for circuits with 30 qubits and fewer, the best runtimes {\em on a laptop} are attained with the basic Schr\"odinger capability\ of our simulator. However, the Microsoft QDK simulator required more than the 16 GiB memory available (Rollright used a little over 8 GiB). We therefore installed Microsoft QDK on an 18-core Linux server with sufficient memory and observed that it used all 72 available threads. 
We first configured Rollright in the Schr\"odinger mode.  As shown in Table \ref{tab:MSFT}, Rollright is $19\times$ faster than the QDK simulator with a $3\times$ memory advantage. The ratios remain similar for circuits of larger depth, and the memory ratio does not depend on how many threads Rollright uses. We then evaluated Rollright in the Schr\"odinger-Feynman mode using the even (30=15+15) cut, with 18 parallel processes and four threads per process.
Rollright was $31\times$ faster and used $886\times$ less memory. The memory ratio depends linearly on how many simultaneous processes Rollright uses. The runtime ratio grows exponentially with circuit depth (specifically, with the number of xCZ gates). Detailed logs of Rollright simulations
are provided in Appendix \ref{app:30q}.

While Rollright significantly outperforms the Microsoft SDK simulator in our comparisons, we consider only quantum-supremacy benchmarks, whereas Microsoft developers may have had different priorities and continue improving their software. Our algorithmic contributions can be useful when simulating many families of quantum circuits, but the technical effort necessary to demonstrate such applications is beyond the scope of this paper. Moreover, Microsoft SDK offers value in areas unrelated to circuit simulation.

\begin{table}[tb]
\begin{center}
\begin{tabular}{|c||r|r||r|r|c|r|r||r|r|}
\hline
Circuit & \multicolumn{2}{|c||}{Gates} & \multicolumn{2}{|c|}{Microsoft QDK} & \multicolumn{3}{|c||}{U. of Michigan Rollright}
 & \multicolumn{2}{|c|}{QDK/Rollright} \\
 \small
depth 1+26+1 & All & 2-q & Time (s) & Memory (MiB) & Mode & Time & Memory & Time & Memory \\
\hline
\multicolumn{10}{c}{{\bf MacBook Pro 2017} --- MacOS High Sierra: 16 GiB, Intel Core i7-7700HQ (2.80GHz) 4 cores 8 threads} \\
\hline \small
\verb/inst_4x4_27_5/ & 274 & 78 & 2.44 & --- & S & $<0.1$& --- & --- & --- \\
\hline \small
\verb/inst_4x6_27_0/ & 417 & 123 & 14.22 & 452 & S & 1.19 & 128 & 11.95 & 3.53 \\
\hline \small
\verb/inst_5x5_27_5/ & 435 & 130 & 27.47 & 837 & S & 1.68 & 256 & 16.35 & 3.27 \\
\hline \small
\verb/inst_5x6_27_5/ & 524 & 161 & --- & out of memory & S & 72 & 8192 & --- & --- \\
\hline
\multicolumn{10}{c}{{\bf Server} --- Ubuntu Linux: 144 GiB, Intel Xeon Platinum 8124M (3.00GHz) 18 cores, 72 threads} \\
\hline \small
\verb/inst_5x6_27_5/ & 524 & 161 & 400.72 & 23930.58 & \bf S & 24.4 & 8192 & \bf 16.42 & \bf  2.92 \\
\hline \small
\verb/inst_5x6_27_5/ & 524 & 161 & 400.72 & 23930.58 & \bf S-F & 12.95 & 27.00 & \bf 30.94 & \bf 886.32 \\
\hline
\end{tabular}
\parbox{15cm}{
\caption{\label{tab:MSFT} Comparisons of our simulator Rollright to the simulator from Microsoft QDK v0.2.1806.3001 (June 30 2018)
on 24-, 25- and 30-qubit benchmarks from Google (v2)  from Section~\ref{sec:bench}, performed on 
a laptop and a mid-range server. Memory usage is the increase in max resident memory versus the baseline 16-qubit simulation, to exclude code-segment size. In 30-qubit simulation, Rollright attains greater improvements (over Microsoft QDK) in the Schr\"odinger-Feynman mode with 18 parallel processes than in the single-process Schr\"odinger mode.
}}
\end{center}
\vspace{-4mm}
\end{table}

\section{Comparisons to IBM QISKit and IBM Q}\label{app:QASM}

IBM QISKit ({\tt https://qiskit.org/}) is an open-source quantum computing framework, which includes several simulators in the software toolchain. We used QISKit-Terra v0.5.7 (July 19 2018) and followed the comparison protocol in Appendix~\ref{app:QDK}. We found the QASM simulator to be faster than the state-vector simulator on quantum-supremacy circuits. Comparisons of our simulator Rollright to QASM 
are reported in Table \ref{tab:IBM} and follow the trends in comparisons to Microsoft QDK 
in Table \ref{tab:MSFT}. On a depth 1+26+1 30-qubit circuit, Rollright runs $16\times$ faster and uses
$590\times$ less memory than QASM. On a depth 1+26+1 32-qubit circuit, Rollright runs $430.6\times$
faster and uses $1331\times$ less memory than QASM, even though QASM exhibits a high degree 
of parallelism with 6010\% CPU utilization rate. The 32-qubit circuit uses the oblong $4x8$ 
qubit array, and the Schr\"odinger-Feynman mode of our simulator is able to exploit this shape.
Therefore, we consider results for both the $5x6$ and the $4x8$ qubit arrays important.
Detailed logs of Rollright simulations are provided in Appendix \ref{app:30q}.

QISKit-Terra offers the option to run simulations on remote servers
with Power9 CPUs and NVIDIA GPUs through the IBM Q service. The last four rows of Table \ref{tab:IBM} compare such simulations with single-server CPU-based Rollright simulations in two modes, with considerable wins for Rollright. Enabling GPU support in the IBM simulator does not impact memory usage (we do not measure memory usage inside the GPU), but improves runtime by almost 30\% on 30-qubit circuits.
However, the GPU-based version is twice as slow on 32-qubit circuits. This may be due to the limitations
of GPUs in terms of on-board memory and memory bandwidth, in which case future generations of GPUs
may be helpful. In any case, Rollright exhibits a $430.6\times$ speedup and $1331\times$ memory advantage for 32-qubit circuits, likely beyond upcoming GPU improvements.

We also compared the performance of QISKit on IBM Q to that of Rollright on a MacBook Pro. Even in such an unfair comparison, Rollright is twice as fast, while using 30-70\% less memory.
Implementing our algorithms on GPUs is likely to give a significant additional speedup. 
On the other hand, modern GPU resources can be costly. Whether they improve
the total cost of simulation remains an open question.

 \begin{table}[tb]
\begin{center}
\begin{tabular}{|c||r|r||r|r|c|r|r||r|r|}
\hline
Circuit & \multicolumn{2}{|c||}{Gates} & \multicolumn{2}{|c|}{IBM QISKit (QASM) } & \multicolumn{3}{|c||}{U. of Michigan Rollright}
 & \multicolumn{2}{|c|}{QISKit/Rollright} \\
 \small 
 depth 1+26+1 & All & 2-q & Time (s) & Memory (MiB) & Mode & Time & Memory & Time & Memory \\
\hline
\multicolumn{10}{c}{{\bf MacBook Pro 2017} --- MacOS High Sierra: 16 GiB, Intel Core i7-7700HQ (2.80GHz) 4 cores 8 threads} \\
\hline \small
\verb/inst_4x4_27_5/ & 274 & 78 & 1.59 & --- & S & $<0.1$& --- & --- & --- \\
\hline \small
\verb/inst_4x6_27_0/ & 417 & 123 & 9.87 & 176.34 & S & 1.19 & 128 & 8.29 & 1.38 \\
\hline \small
\verb/inst_5x5_27_5/ & 435 & 130 & 18.3 & 432.36 & S & 1.68 & 256 & 10.89 & 1.69 \\
\hline \small
\verb/inst_5x6_27_5/ & 524 & 161 & --- & out of memory & S & 72 & 8192 & --- & --- \\
\hline
\multicolumn{10}{c}{{\bf Local server} --- Ubuntu Linux: 144 GiB, Intel Xeon Platinum 8124M (3.00GHz) 18 cores, 72 threads} \\
\hline \small
\verb/inst_5x6_27_5/ & 212 & 161 & 212.25 & 15925 & \bf S & 24.4 & 8192 & \bf 8.70 & \bf  1.94 \\
\hline \small
\verb/inst_5x6_27_5/ & 212 & 161 & 212.25 & 15925 & \bf S-F & 12.95 & 27.00 & \bf 16.39 & \bf 589.83 \\
\hline \small
\verb/inst_4x8_27_5/ & 560 & 168 & 503.8 & 63922 & \bf S & 129 & 32000 & \bf 3.91 & \bf  2.00 \\
\hline \small
\verb/inst_4x8_27_5/ & 560 & 168 & 503.8 & 63922 & \bf S-F & 1.17 & 48.00 & \bf 430.60 & \bf 1331.70 
\\
\hline
\multicolumn{10}{c}{{\bf Server} --- QISKit-Terra on IBM Power9 with NVIDIA GPUs vs. Rollright on Intel Xeon 72 threads} \\
\hline \small
\verb/inst_5x6_27_5/ & 524 & 161 & 165.66 & 15925 & \bf S & 24.4 & 8192 & \bf 6.79 & \bf  1.94 \\
\hline \small
\verb/inst_5x6_27_5/ & 524 & 161 & 165.66 & 15925 & \bf S-F & 12.95 & 27.00 & \bf 12.79 & \bf 589.83 \\
\hline
\hline \small
\verb/inst_4x8_27_5/ & 560 & 168 & 1026.28 & 63922 & \bf S & 129 & 32000 & \bf 7.96 & \bf  2.00 \\
\hline \small
\verb/inst_4x8_27_5/ & 560 & 168 & 1026.28 & 63922 & \bf S-F & 1.17 & 48.00 & \bf 877.16 & \bf 1331.70 
\\
\hline
\end{tabular}
\parbox{15cm}{
\caption{\label{tab:IBM} Comparisons of our simulator Rollright  to the QASM simulator from IBM QISKit-Terra v0.5.7 (July 19 2018) 
on 24-, 25- and 30-qubit Google v2 benchmarks from Section~\ref{sec:bench}, performed on a laptop and a mid-range server. Memory usage is  the increase in max resident memory versus the baseline 16-qubit simulation, to exclude code-segment size. The trends mirror those in
Table \ref{tab:MSFT}. The last two rows show results of GPU-based simulation using QISKit with the IBM Q service. 
}}
\end{center}
\vspace{-4mm}
\end{table}

\section{Validation of simulation results} 
\label{app:validation}

For Google quantum-supremacy circuits, simulation results can be checked using approximate simulation. Consider the correct final state $|\psi\rangle$ of simulation and a norm-1 approximate state $|\psi_a\rangle$ with $|\langle\psi|\psi_a\rangle|^2 =f$. As shown above,
multiple $|\psi_a\rangle$ states can be produced $f$ times faster (each) than the $|\psi\rangle$ state.
Such approximate states help validate claims of producing $|\psi\rangle$. In the following two-party validation protocol, we assume a quantum circuit that the Verifier can simulate to fidelity $f<1$
in the sense that the Verifier can approximate the final simulation state with fidelity $f$ and do so with
a sufficient variety of states at random. For each state, the Verifier and the Claimant save $k$ amplitudes
so as to estimate fidelities between their states to accuracy $\delta \le f$.
\begin{enumerate}
 \item The Verifier gives the Claimant $k$ pseudo-randomly generated amplitude indices
          (we used $k$=1M and specified the indices by a PRNG seed).
 \item The Claimant simulates the circuit and saves $k$ amplitudes with given indices.
 \item The Verifier picks a fidelity value $0 \ll f_1 < 1 $ and simulates the circuit with fidelity $f_1$
         producing one of many possible $| \psi_a \rangle$ states and saving $k$ amplitudes with the specified indices.
 \item The Claimant shares their amplitudes with the Verifier.
 \item The Verifier estimates fidelity $f_e$ between the Verifier's and the Claimant's states.
 \item If $|f_e - f_1| > \delta$, the test fails. Else, the test may be repeated to increase
         the certainty of passing.
\end{enumerate}
A handful of sufficiently unpredictable approximate states
suffice to validate $|\psi\rangle$ because for a generic approximate state with fidelity $f_1 \ll 1$,
the expected state fidelity w.r.t $|\psi_a\rangle$ is $f_1 f \ll f$. Moreover, the Verifier can estimate the fidelity
of the Claimant's state. The Verifier can select circuits from some family and repeat the test for different circuits.

\section{30- and 32-qubit simulations}
\label{app:30q}

   Below we include detailed simulation logs in support of comparisons to Microsoft and IBM simulators
   in Appendices~\ref{app:QDK} and \ref{app:QASM}.

\subsection{30-qubit depth $1+26+1$ simulation on a laptop}
This Schr\"odinger-mode simulation uses the circuit \verb+inst_5x6_27_5+ publicly available from \cite{new_benchmarks}.
\begin{verbatim}
(C) 2017, 2018  Regents of the University of Michigan
Rollright ver 2.3 - a quantum circuit simulator
Igor L. Markov and Aneeqa Fatima

Hostname : Aneeqa-Macbook Darwin Kernel Version 17.4.0: Sun Dec 17 09:19:54 PST 2017
CPU model name : Intel(R) Core(TM) i7-7700HQ CPU @ 2.80GHz
CPU cores : 4
Hardware threads : 8
Max threads per process : 8
L2 cache size : 262144
L3 cache size : 6291456
CPU instructions width : popcnt:4, sse4.2:256, avx:512, avx2:1024
Using instructions : AVX-2, popcnt

Compiler : gcc 7.3.0
Compiled on : Jul 24 2018 02:04:06
Executed on : 7/27/2018 11:0:37
Size of complex : 8 B

Verbosity : 3
Circuit file : inst_5x6_27_5.txt
Circuit type : Google
Qubits : 30  Gates : 524 (161 two q gates) Cycles : 28
Simulation type : full state-vector  
Low-value qubits : 15 q
Layers simulated : H (2), CZ & T (13), X & Y & H (13)

State representation size : 8 GiB 
Norm : 1
Mean entropy : 22 Cross entropy : 23.2
Probabilities : 2.18e+23(min), 8.64e+33(max), 9.31e-10(avg)
Log_2 (max / min) = 35.2
Avg inaccuracy per probability > 5.89e-17 (6.32e-06%)

amp[3]  	= -1.96742e-05 + 1.59778e-05j
amp[1/4]	= 3.55476e-08 + 4.40849e-06j
amp[1/2]	= -1.06781e-05 + 1.58316e-06j
amp[3/4]	= 1.256e-06 + 4.03744e-05j
amp[-3] 	= 3.07262e-05 - 3.1141e-05j

Runtime (71 s total) by category 
     Initial H (30)                                : 4.06 s    =  5.72%
     Last H (12)                                   : 5.16 s    =  7.27%
     CZ & T (280), Low X & Y (90) & H (14)         : 21.4 s    =  30.1%
     High X & Y (94) & H (4)                       : 39.5 s    =  55.7%
     Rescaling passes (1)                          : 0.881 s   =  1.24%
     Storing amps                                  : 0.0023 s  =  0.00324%
                                                                  -------
     Total                                                        94.3%

Average time per gate : 0.135 s
\end{verbatim}

\subsection{30-qubit depth $1+26+1$ simulation on a server}
This Schr\"odinger-mode simulation uses the circuit \verb+inst_5x6_27_5+ publicly available from \cite{new_benchmarks}.
\begin{verbatim}
(C) 2017, 2018  Regents of the University of Michigan
Rollright ver 2.3 - a quantum circuit simulator
Igor L. Markov and Aneeqa Fatima

Hostname : Linux ip-172-31-24-65 4.4.0-1049 #58-Ubuntu SMP Fri Jan 12 23:17:09 UTC 2018
CPU model name	: Intel(R) Xeon(R) Platinum 8124M CPU @ 3.00GHz
cpu cores	: 18
Hardware threads : 72
Max threads per process : 72
MemTotal:       144156168 kB
L3 cache size	: 25344 KB
CPU instructions width : popcnt:4, sse4.2:256, avx:512, avx2:1024
Using instructions : AVX-2, popcnt

Compiler : gcc 7.2.0
Compiled on : Jul 27 2018 06:58:56
Executed on : 7/27/2018 14:59:50
Size of complex : 8 B

Verbosity : 3
Circuit file : inst_5x6_27_5.txt
Circuit type : Google
Qubits : 30  Gates : 524 (161 two q gates) Cycles : 28
Simulation type : full state-vector  
Low-value qubits : 15 q
Layers simulated : H (2), CZ & T (13), X & Y & H (13)

State representation size : 8 GiB 
Norm : 1
Mean entropy : 22 Cross entropy : 23.2
Probabilities : 2.18e+23(min), 8.64e+33(max), 9.31e-10(avg)
Log_2 (max / min) = 35.2
Avg inaccuracy per probability > 5.89e-17 (6.32e-06%)

amp[3]  	= -1.96742e-05 + 1.59778e-05j
amp[1/4]	= 3.55476e-08 + 4.40849e-06j
amp[1/2]	= -1.06781e-05 + 1.58316e-06j
amp[3/4]	= 1.256e-06 + 4.03744e-05j
amp[-3] 	= 3.07262e-05 - 3.1141e-05j

Runtime (24.4 s total) by category 
	Initial H (30)                          : 0.432 s    =  1.77%
	Last H (12)                             : 1.85 s     =  7.6%
	CZ & T (280), Low X & Y (90) & H (14)   : 3.51 s     =  14.4%
	High X & Y (94) & H (4)                 : 18.4 s     =  75.3%
	Rescaling passes (1)                    : 0.226 s    =  0.925%
	Storing amps                            : 3.65e-05 s =  0.00015%
                                                        -------
Total                                                   98.2%

Average time per gate : 0.0465 s
\end{verbatim}

\subsection{30-qubit depth $1+26+1$ cloud simulation}
This Schr\"odinger-Feynman simulation uses the circuit \verb+inst_5x6_27_5+ publicly available from \cite{new_benchmarks}.
\begin{verbatim}
(C) 2017, 2018  Regents of the University of Michigan
Rollright ver 2.3 - a quantum circuit simulator
Igor L. Markov and Aneeqa Fatima

Hostname : Linux ip-172-31-24-65 #58-Ubuntu SMP Fri Jan 12 23:17:09
CPU model name	: Intel(R) Xeon(R) Platinum 8124M CPU @ 3.00GHz
cpu cores	: 18
Hardware threads : 72
Max threads per process : 4
MemTotal:       144156168 kB
L3 cache size	: 25344 KB
CPU instructions width : popcnt:4, sse4.2:256, avx:512, avx2:1024
Using instructions : AVX-2, popcnt

Compiler : gcc 7.2.0
Compiled on : Jul 27 2018 06:58:56
Executed on : 7/27/2018 15:26:49
Size of complex : 8 B
Verbosity : 3

Circuit file : inst_5x6_27_5.txt
Circuit type : Google
Qubits : 30  Gates : 524 (161 two q gates) Cycles : 28
Qubit grid 5x6 with 2 blocks 
     0^   1^   2^   3_   4_   5_  
     6^   7^   8^   9_  10_  11_  
    12^  13^  14^  15_  16_  17_  
    18^  19^  20^  21_  22_  23_  
    24^  25^  26^  27_  28_  29_  
 Cross-gates : 5

Simulation type : sum of tensor products / single cut
Cut : vertical 15q + 15q (17 xCZ gates)
Simulating xCZ gates : using projection-based branches
xCZ path breakdown : 10p + 5r + 2b (cycle breakdown : 19p + 8r + 1b)
Low-value qubits : 8 q, 8 q
Requested num amps : 1000

Multi-process simulation : 
    1024 processes (4 threads each) over 1 node in 18 batches per node 
    State representation size : 512 KiB 
    Peak memory : 27648.0 KiB
    Layers simulated : H (193), CZ & T (330), X & Y & H (330)
    Layers breakdown : 11p + 192r + 128b
     Batch stats :
          Avg user time : 44.429 s 
          Wallclock : 12.725 s (avg), 12.91 s (max)
          Avg CPU utilization : 348.36% (87.09% per thread)
          Avg resident size : 5.816 MiB
          Avg page faults : 0.018 (major), 5795.459 (minor)
     Billable runtime : 3.586e-03 hrs (3.586e-06 hrs per amp)

amp[3]  	= -1.967417e-05+1.597779e-05j
amp[1/4]	= 3.556267e-08+4.408506e-06j
amp[1/2]	= -1.067803e-05+1.583226e-06j
amp[3/4]	= 1.256007e-06+4.037441e-05j
amp[-3] 	= 3.072615e-05-3.114105e-05j

Avg runtime (0.2 s total) per process by category 
    Initial H (30)                        : 0.00328 s    = 1.636%
    Last H (14)                           : 0.007 s      = 3.391%
    xCZ (17)                              : 0.009 s      = 4.505%
    CZ & T (263), Low X & Y (84) & H (12)	: 0.038 s      = 19.066%
    Single X (9) & Y (11)                 : 0.008 s      = 4.021%
    High X & Y (80) & H (4)               : 0.007 s      = 3.51%
    Rescaling passes (32)                 : 0.001 s      = 0.355%
    Copying (127)                         : 0.006 s      = 3.071%
    Storing amps                          : 0.047 s      = 23.453%
                                                           ----------
    Total                                                  63.007%

Avg time per gate : 3e-06 s
Per-process runtime breakdown : 
     Prefix             : 0.068 s = 34.159%
     Branches           : 0.137 s = 68.47%
     Memory mapped I/O  : 2e-05 s = 0.00799%
\end{verbatim}

\subsection{32-qubit depth $1+26+1$ simulation on a server}
This Schr\"odinger-mode simulation uses the circuit \verb+inst_4x8_27_0+ publicly available from \cite{new_benchmarks}.
\begin{verbatim}
C) 2017, 2018  Regents of the University of Michigan
Rollright ver 2.3 - a quantum circuit simulator
Igor L. Markov and Aneeqa Fatima

Hostname : Linux ip-172-31-90-201 #58-Ubuntu SMP Fri Jan 12 23:17:09 
CPU model name	: Intel(R) Xeon(R) Platinum 8124M CPU @ 3.00GHz
cpu cores	: 18
Hardware threads : 72
Max threads per process : 72
MemTotal:       144156168 kB
L3 cache size	: 25344 KB
Hugepagesize:       2048 kB
HugePages_Total:       0
HugePages_Free:        0
CPU instructions width : popcnt:4, sse4.2:256, avx:512, avx2:1024
Using instructions : AVX-2, popcnt

Compiler : gcc 7.3.0
Compiled on : Sep 17 2018 07:21:48
Executed on : 9/17/2018 7:51:31
Size of complex : 8 B

Verbosity : 3
Circuit file : inst_8x4_27_0.txt
Circuit type : Google
Qubits : 32  Gates : 560 (168 two-q gates) Cycles : 28
Simulation type : full state-vector  
Low-value qubits : 16 q
Layers simulated : H (2), CZ & T (13), X & Y & H (13)

State representation size : 32 GiB 
Norm : 1
Mean entropy : 23.5 Cross entropy : 24.6
Probabilities : 7.29e-07(min), 9.39e+04(max), 2.33e-10(avg)
Log_2 (max / min) = 36.9
Avg inaccuracy per probability > 3.27e-17 (1.41e-05%)

Correctness check : no data available
amp[3]  	= -3.43324e-06 - 3.52942e-06j
amp[1/4]	= 1.19572e-05 + 6.78437e-06j
amp[1/2]	= 1.11328e-05 - 2.94045e-06j
amp[3/4]	= -7.43047e-06 - 2.17285e-06j
amp[-3] 	= 9.70701e-06 + 7.38354e-06j

Runtime (129 s total) by category 
     Initial H (32)                          : 1.29 s  		   =  1%
     Last H (10)                             : 5.78 s  		   =  4.47%
     CZ & T (298), Low X & Y (90) & H (16)   : 18.8 s  		   =  14.6%
     Single X (2) & Y (4)                    : 16 s  		     =  12.3%
     High X & Y (102) & H (6)                : 85.3 s  		   =  66%
     Rescaling passes (2)                    : 2.14 s		     =  1.66%
     Storing amps                            : 3.36e-05 s  =  2.6e-05%
                                                              -------
     Total                                                    99%

Average time per gate : 0.231 s
\end{verbatim}

\subsection{32-qubit depth $1+26+1$ cloud simulation}
This Schr\"odinger-Feynman simulation uses the circuit \verb+inst_8x4_27_0+ publicly available from \cite{new_benchmarks}.
\begin{verbatim}
(C) 2017, 2018  Regents of the University of Michigan
Rollright ver 2.3 - a quantum circuit simulator
Igor L. Markov and Aneeqa Fatima

Hostname : Linux ip-172-31-90-201 #58-Ubuntu SMP Fri Jan 12 23:17:09
CPU model name	: Intel(R) Xeon(R) Platinum 8124M CPU @ 3.00GHz
cpu cores	: 18
Hardware threads : 72
Max threads per process : 4
MemTotal:       144156168 kB
L3 cache size	: 25344 KB
Hugepagesize:       2048 kB
HugePages_Total:       0
HugePages_Free:        0
CPU instructions width : popcnt:4, sse4.2:256, avx:512, avx2:1024
Using instructions : AVX-2, popcnt

Compiler : gcc 7.3.0
Compiled on : Sep 17 2018 07:21:48
Executed on : 9/17/2018 7:44:17
Size of complex : 8 B

Verbosity : 3
Circuit file : inst_8x4_27_0.txt
Circuit type : Google
Qubits : 32  Gates : 560 (168 two-q gates) Cycles : 28
Qubit grid 4x8 with 2 blocks 
     0^   1^   2^   3^   4_   5_   6_   7_  
     8^   9^  10^  11^  12_  13_  14_  15_  
    16^  17^  18^  19^  20_  21_  22_  23_  
    24^  25^  26^  27^  28_  29_  30_  31_  
 Cross-gates : 4

Simulation type : sum of tensor products / single cut
Cut : vertical 16q + 16q (12 xCZ gates)
Simulating xCZ gates : using projection-based branches
xCZ path breakdown : 5p + 5r + 2b (cycle breakdown : 13p + 12r + 3b)
Low-value qubits : 8 q, 8 q
Requested num amps : 1000

Multi-process simulation : 
     32 processes (4 threads each) over 1 node in 16 batches per node 
     State representation size : 1 MiB 
     Peak memory : 48.0 MiB
     Layers simulated : H (129), CZ & T (519), X & Y & H (519)
     Layers breakdown : 8p + 256r + 256b
     Batch stats :
          Avg user time : 3.834 s
          Wallclock : 1.031 s (avg), 1.17 s (max)
          Avg CPU utilization : 372.5% (93.125% per thread)
          Avg resident size : 7.242 MiB
          Avg page faults : 0.5 (major), 9428.438 (minor)
     Billable runtime : 3.250e-04 hrs (3.250e-07 hrs per amp)

amp[3]  	= -3.433250e-06-3.529406e-06j
amp[1/4]	= 1.195717e-05+6.784386e-06j
amp[1/2]	= 1.113274e-05-2.940453e-06j
amp[3/4]	= -7.430470e-06-2.172844e-06j
amp[-3] 	= 9.707015e-06+7.383564e-06j

Avg runtime (0.489 s total) per process by category 
     Initial H (32)                         : 0.00673 s = 1.376%
     Last H (12)                            : 0.01 s  	 = 1.951%
     xCZ (12)                               : 0.015 s  	= 3.097%
     CZ & T (286), Low X & Y (84) & H (12)  : 0.149 s  	= 30.423%
     Single X (8) & Y (10)                  : 0.024 s  	= 4.854%
     High X & Y (96) & H (8)                : 0.051 s  	= 10.51%
     Rescaling passes (32)	                 : 0.001 s  	= 0.292%
     Copying (127)                          : 0.012 s  	= 2.424%
     Storing amps                           : 0.137 s  	= 28.071%
                                                          ----------
     Total                                                82.997%

Avg time per gate : 7e-06 s
Simulation per process runtime breakdown : 
     Prefix             : 0.143 s = 29.158%
     Branches           : 0.356 s = 72.9%
     Memory-mapped I/O  : 1e-05 s = 0.00219%
\end{verbatim}

\section{ ``Easy'' 42-qubit depth $1+25+1$ simulation}
This Schr\"odinger-Feynman simulation uses the circuit \verb+inst_7x6_26_0+ publicly available from \cite{new_benchmarks}, Such revised benchmarks  
are harder to simulate than older quantum-supremacy benchmarks
(see Section~\ref{sec:bench}).

\begin{verbatim}

(C) 2017, 2018  Regents of the University of Michigan
Rollright ver 2.3 - a quantum circuit simulator
Igor L. Markov and Aneeqa Fatima

Hostname : Linux boixo-rr-h96-5 4.13.0-45-generic #50-Ubuntu
CPU model name	: Intel(R) Xeon(R) CPU @ 2.00GHz
cpu cores	: 24
Hardware threads : 96
Max threads per process : 4
MemTotal:       89073048 kB
L3 cache size	: 56320 KB
Hugepagesize:       2048 kB
HugePages_Total:       0
HugePages_Free:        0
CPU instructions width : popcnt:4, sse4.2:256, avx:512, avx2:1024
Using instructions : AVX-2, popcnt

Compiler : gcc 7.2.0
Compiled on : Jul 24 2018 02:36:06
Executed on : 7/29/2018 9:16:36
Size of complex : 8 B

Verbosity : 3
Circuit file : inst_7x6_26_0
Circuit type : Google
Qubits : 42  Gates : 711 (224 two q gates) Cycles : 27
Qubit grid 7x6 with 2 blocks 
     0^   1^   2^   3^   4^   5^  
     6^   7^   8^   9^  10^  11^  
    12^  13^  14^  15^  16^  17^  
    18^  19^  20^  21^  22^  23^  
    24_  25_  26_  27_  28_  29_  
    30_  31_  32_  33_  34_  35_  
    36_  37_  38_  39_  40_  41_  
 Cross-gates : 6

Simulation type : sum of tensor products / single cut
Cut : horizontal 24q + 18q (18 xCZ gates)
Simulating xCZ gates : using projection-based branches
xCZ path breakdown : 6p + 6r + 6b (cycle breakdown : 13p + 8r + 6b)
Low-value qubits : 12 q, 9 q
Requested num amps : 1000000

Multi-process simulation : 
     64 processes (4 threads each) over 1 node in 23 batches per node 
     State representation size : 130 MiB 
     Peak memory : 8970.0 MiB
     Layers simulated : H (4097), CZ & T (16775), X & Y & H (12679)
     Layers breakdown : 8p + 384r + 16384b
     Batch stats :
          Avg user time : 10.26 hrs 
          Wallclock : 5.534 hrs (avg), 6.016 hrs (max)
          Avg CPU utilization : 185.766% (46.441% per thread)
          Avg resident size : 435.466 MiB
          Avg page faults : 0.406 (major), 14708868.438 (minor)
     Billable runtime : 6.016e+00 hrs (6.016e-06 hrs per amp)
     Avg zero count 
          1st Checkpoint : A = 0 (0.0%), B = 0 (0.0%)
          2nd Checkpoint : A = 0 (0.0%), B = 0 (0.0%)

amp[3]    = 2.998598e-07-2.862689e-07j
amp[1/4]  = 2.248877e-07+1.877150e-07j
amp[1/2]  = 2.425415e-07-1.685559e-08j
amp[3/4]  = -4.233980e-07-7.303483e-08j
amp[-3]   = 4.674360e-07-2.024549e-07j

Avg runtime (7147.031 s total) per process by category 
     Initial H (42)                          : 0.03196 s     = 0.0%
     Last H (22)                             : 411.156 s     = 5.753%
     xCZ (18)                                : 168.734 s     = 2.361%
     CZ & T (363), Low X & Y (108) & H (20)  : 917.344 s     = 12.835%
     Single X (7) & Y (5)                    : 95.763 s      = 1.34%
     High X & Y (126)                        : 635.188 s     = 8.887%
     Rescaling passes (64)                   : 1.199 s       = 0.017%
     Copying (4095)                          : 131.828 s     = 1.845%
     Storing amps                            : 4620.312 s    = 64.647%
                                                               ----------
     Total                                                     97.685%

Avg time per gate : 0.002454 s
Simulation per process runtime breakdown : 
     Prefix               : 175.703 s = 2.458%
     Branches             : 7100.469 s = 99.349%
     Memory mapped I/O    : 0.01171 s = 0.00016%


\end{verbatim}

\section{ 64-qubit depth $1+22$ simulation}\label{app:64_22}

Recent work related to ours \cite{64q} used GPUs and a distributed cluster with 128 nodes to simulate quantum circuits. The layout of CZ two-qubit gates used in that work was changed relative to the prior work to lower the cost of classical simulation by approximately 16 times, which prevents a direct comparison of runtimes in general: we focus instead on the circuits from Ref.~\cite{new_benchmarks}, which are significantly harder to simulate. Nevertheless, we performed a simulation of a circuit with $8 \times 8$ qubits and depth $1+22$ with the choice for the layout of CZs from Ref.~\cite{64q}.  Our simulation took 18.6 hours on 64 {\tt n1-highmem-32} Google Cloud virtual machines, a 1.7$\times$ runtime improvement using half as many nodes and running in a commercial cloud environment.\footnote{The nodes in the simulation of Ref.~\cite{64q} have each a Xeon E5-2690V4 processor with 14 cores. We use {\tt n1-highmem-32} virtual machines in Google Cloud, with 32 vCPUs or hyper-threads, equivalent to 16 cores 2.0 GHz Intel (Skylake) platform.}

\begin{verbatim}
(C) 2017, 2018  Regents of the University of Michigan
Rollright ver 2.3 - a quantum circuit simulator
Igor L. Markov and Aneeqa Fatima

Hostname : Linux boixo-rr-hmm32-36 4.13.0-45-generic #50-Ubuntu
CPU model name	: Intel(R) Xeon(R) CPU @ 2.00GHz
CPU cores	: 16
Hardware threads : 32
Max threads per process : 32
MemTotal:       214601516 kB
L3 cache size	: 56320 KB
CPU instructions width : popcnt:4, sse4.2:256, avx:512, avx2:1024
Using instructions : AVX-2, popcnt

Compiler : gcc 7.2.0
Compiled on : Jun 29 2018 21:51:16
Executed on : 6/30/2018 18:20:54
Size of complex : 8 B

Verbosity : 3
Circuit file : inst_8_8_22_0_chen
Circuit type : Google
Qubits : 64  Gates : 848  Cycles : 23
Qubit grid 8x8 with 2 blocks
     0^   1^   2^   3^   4^   5^   6^   7^
     8^   9^  10^  11^  12^  13^  14^  15^
    16^  17^  18^  19^  20^  21^  22^  23^
    24^  25^  26^  27^  28^  29^  30^  31^
    32_  33_  34_  35_  36_  37_  38_  39_
    40_  41_  42_  43_  44_  45_  46_  47_
    48_  49_  50_  51_  52_  53_  54_  55_
    56_  57_  58_  59_  60_  61_  62_  63_
 Cross-gates : 8

Simulation type : sum of tensor products / single cut
Cut : horizontal 32q + 32q (16 xCZ gates)
Simulating xCZ gates : using projection-based branches
xCZ path breakdown : 8p + 4r + 4b (cycle breakdown : 16p + 0r + 7b)
Low-value qubits : 16 q, 16 q
Requested num amps : 16384

Multi-process simulation :
     256 processes (32 threads each) over 64 nodes in 1 batch per node
     State representation size : 64 GiB
     Peak memory : 12288.0 GiB (192.0 GiB per node)
     Layers simulated : H (1), CZ & T (1048), X & Y (1048)
     Layers breakdown : 9p + 16r + 1024b
     Batch stats :
          Avg user time : 254.053 hrs
          Wallclock : 14.726 hrs (avg), 18.593 hrs (max)
          Avg CPU utilization : 1734.633% (54.207% per thread)
          Avg resident size : 187.505 GiB
          Avg page faults : 0.0 (major), 301993529.113 (minor)
     Billable runtime : 1.190e+03 hrs (7.263e-02 hrs per amp)

amp[3]  	= -3.572132e-11-1.877656e-10j
amp[1/4]	= 2.245370e-10+2.368543e-10j
amp[1/2]	= 2.019670e-10+1.266517e-10j
amp[3/4]	= 7.522537e-11+8.860908e-11j
amp[-3] 	= -1.860613e-10-2.177499e-10j

Avg runtime (13207.812 s total) per process by category
     Initial H (64)                 : 1.8916 s  	  = 0.014%
     xCZ (16)                       : 454.203 s  	 = 3.439%
     CZ & T (480), Low X & Y (138)	 : 3210.234 s  	= 24.306%
     Single X (7) & Y (3)           : 2404.062 s  	= 18.202%
     High X & Y (140)               : 6346.875 s  	= 48.054%
     Rescaling passes (1)           : 1.872 s  	   = 0.014%
     Copying (255)                  : 699.098 s  	 = 5.293%
     Storing amps                   : 6.254 s  	   = 0.047%
                                                     ----------
     Total                                           99.369%

Avg time per gate : 0.060839 s
Simulation per process runtime breakdown :
	Prefix                    : 857.234 s   = 6.49%
	Branches                  : 13005.469 s = 98.468%
	Memory mapped I/O         : 0.00015 s   = 0.0%
	
	
\end{verbatim}

\section{49-qubit large-depth simulations}
Here we report simulations of circuits on $7\times 7$ qubits of depth 1+39+1,
1+40+1 and 1+48+1.

\subsection{ 49-qubit depth $1+39+1$ simulation with fidelity 1\%}
This simulation uses the circuit \verb+inst_7x7_40_0+ publicly available from \cite{new_benchmarks}.

\begin{verbatim}
(C) 2017, 2018  Regents of the University of Michigan
Rollright ver 2.2 - a quantum circuit simulator
Igor L. Markov and Aneeqa Fatima

Hostname : Linux boixo-rr-h32-333 4.13.0-43-generic #48-Ubuntu
CPU model name	: Intel(R) Xeon(R) CPU @ 2.30GHz
CPU cores	: 16
Hardware threads : 32
Max threads per process : 8
MemTotal:       123768660 kB
L3 cache size	: 46080 KB
CPU instructions width : popcnt:4, sse4.2:256, avx:512, avx2:1024
Using instructions : AVX-2, popcnt

Compiler : gcc 7.2.0
Compiled on : May 28 2018 21:20:26
Executed on : 5/29/2018 9:37:17
Size of complex : 8 B

Verbosity : 3
Circuit file : inst_7x7_40_0
Circuit type : Google
Qubits : 49  Gates : 1252  Cycles : 41
Qubit grid 7x7 with 2 blocks
     0^   1^   2^   3^   4^   5^   6^
     7^   8^   9^  10^  11^  12^  13^
    14^  15^  16^  17^  18^  19^  20^
    21^  22^  23^  24^  25^  26^  27^
    28_  29_  30_  31_  32_  33_  34_
    35_  36_  37_  38_  39_  40_  41_
    42_  43_  44_  45_  46_  47_  48_
 Cross-gates : 7

Simulation type : sum of tensor products / single cut
Cut : horizontal 28q + 21q (31 xCZ gates)
Requested end-to-end circuit fidelity: 0.01
Approximation type : pruned xCZ branches
Simulating xCZ gates : using projection-based branches
xCZ path breakdown : 24p + 4r + 3b (cycle breakdown : 33p + 4r + 4b)
Low-value qubits : 14 q, 11 q
Requested num amps : 1000000

Multi-process simulation :
     167773 processes (8 threads each) in 4 batch(es) over 617 node(s)
     State representation size : 2.02 GiB
     Peak memory : 14956.08 GiB (24.24 GiB per node)
     Layers simulated : H (129), CZ & T (465), X & Y & H (337)
     Layers breakdown : 18p + 64r + 384b
     Batch stats :
          Avg user time : 249090.762 s
          Wallclock : 64480.76 s (avg), 126850.07 s (max)
          Avg CPU utilization : 390.411% (48.801% per thread)
          Avg resident size : 5.976 GiB
          Avg page faults : 0.057 (major), 9480745.407 (minor)
     Billable runtime : 2.174e+04 hrs (2.174e-02 hrs per amp)

amp[3]  	= -2.017607e-09-1.396589e-09j
amp[1/4]	= 1.534751e-10-2.899289e-09j
amp[1/2]	= -4.921908e-09+2.984601e-09j
amp[3/4]	= 4.053322e-10-2.020723e-09j
amp[-3] 	= 1.646911e-10-2.048092e-09j

Avg runtime (945.34 s total) per process by category
     Initial H (49)                         : 0.18201 s  	= 0.019%
     Last H (25)                            : 233.173 s  	= 24.665%
     xCZ (31)                               : 29.88 s  	  = 3.161%
     CZ & T (665), Low X & Y (220) & H (24) : 206.859 s  	= 21.882%
     Single X (3) & Y (7)                   : 39.322 s   	= 4.16%
     High X & Y (228)                       : 197.519 s  	= 20.894%
     Rescaling passes (9)                   : 0.367 s     = 0.039%
     Copying (144)	                         : 41.846 s  	 = 4.427%
     Storing amps                           : 184.067 s  	= 19.471%
                                                            ----------
     Total                                                  98.717%

Avg time per gate : 0.005899 s
Simulation per process runtime breakdown :
     Prefix             : 115.664 s = 12.235%
     Branches           : 826.921 s = 87.473%
     Memory mapped I/O  : 0.00968 s = 0.00102%


\end{verbatim}

\subsection{ 49-qubit depth $1+40+1$ simulation with fidelity 0.51\%}
This simulation uses the circuit \verb+inst_7x7_41_0+ publicly available from \cite{new_benchmarks}.

\begin{verbatim}
(C) 2017, 2018  Regents of the University of Michigan
Rollright ver 2.2 - a quantum circuit simulator
Igor L. Markov and Aneeqa Fatima

Hostname : Linux boixo-rr-hc32-28 4.13.0-43-generic #48-Ubuntu
CPU model name	: Intel(R) Xeon(R) CPU @ 2.00GHz
CPU cores	: 16
Hardware threads : 32
Max threads per process : 8
MemTotal:       29631356 kB
L3 cache size	: 56320 KB
CPU instructions width : popcnt:4, sse4.2:256, avx:512, avx2:1024
Using instructions : AVX-2, popcnt

Compiler : gcc 7.2.0
Compiled on : May 28 2018 21:20:26
Executed on : 6/2/2018 21:21:34
Size of complex : 8 B

Verbosity : 3
Circuit file : inst_7x7_41_0
Circuit type : Google
Qubits : 49  Gates : 1280  Cycles : 42
Qubit grid 7x7 with 2 blocks
     0^   1^   2^   3^   4^   5^   6^
     7^   8^   9^  10^  11^  12^  13^
    14^  15^  16^  17^  18^  19^  20^
    21^  22^  23^  24^  25^  26^  27^
    28_  29_  30_  31_  32_  33_  34_
    35_  36_  37_  38_  39_  40_  41_
    42_  43_  44_  45_  46_  47_  48_
 Cross-gates : 7

Simulation type : sum of tensor products / single cut
Cut : horizontal 28q + 21q (35 xCZ gates)
Requested end-to-end circuit fidelity: 0.0051
Approximation type : pruned xCZ branches
Simulating xCZ gates : using projection-based branches
xCZ path breakdown : 28p + 3r + 4b (cycle breakdown : 37p + 4r + 1b)
Low-value qubits : 14 q, 11 q
Requested num amps : 1000000

Multi-process simulation :
     1369569 processes (8 threads each) in 4 batch(es) over 625 node(s)
     State representation size : 2.02 GiB
     Peak memory : 15150.0 GiB (24.24 GiB per node)
     Layers simulated : H (145), CZ & T (179), X & Y & H (43)
     Layers breakdown : 20p + 32r + 128b
     Batch stats :
          Avg user time : 615850.294 s
          Wallclock : 132293.755 s (avg), 209608.71 s (max)
          Avg CPU utilization : 541.57% (67.696% per thread)
          Avg resident size : 5.975 GiB
          Avg page faults : 0.008 (major), 5286338.665 (minor)
     Billable runtime : 3.639e+04 hrs (3.639e-02 hrs per amp)

amp[3]  	= -2.405457e-09+2.098312e-10j
amp[1/4]	= 3.961961e-10+1.482872e-09j
amp[1/2]	= 6.435293e-11+1.481697e-09j
amp[3/4]	= -7.280204e-10+1.029079e-09j
amp[-3] 	= 1.406264e-09-6.950104e-10j

Avg runtime (238.985 s total) per process by category
     Initial H (49)                         : 0.12541 s     = 0.052%
     Last H (25)                            : 90.339 s      = 37.801%
     xCZ (35)                               : 16.716 s      = 6.995%
     CZ & T (678), Low X & Y (224) & H (18)	: 17.128 s     	= 7.167%
     Single X (6) & Y (5)                   : 3.409 s       = 1.426%
     High X & Y (234) & H (6)               : 20.769 s      = 8.691%
     Rescaling passes (6)                   : 1.121 s       = 0.469%
     Copying (136)                          : 24.359 s      = 10.193%
     Storing amps                           : 56.75 s       = 23.746%
                                                              ----------
     Total                                                    96.541%

Avg time per gate : 0.001459 s
Simulation per process runtime breakdown :
     Prefix            : 54.786 s  = 22.924%
     Branches          : 182.966 s = 76.56%
     Memory mapped I/O : 0.00758 s = 0.00317%


\end{verbatim}

\subsection{49-qubit depth $1+48+1$ simulation with fidelity $2^{-22}$}
This simulation uses the circuit \verb+inst_7x7_49_0+ publicly available from \cite{new_benchmarks}.
\begin{verbatim}

(C) 2017, 2018  Regents of the University of Michigan
Rollright ver 2.3 - a quantum circuit simulator
Igor L. Markov and Aneeqa Fatima

Hostname : Linux boixo-rr-hc32-28 4.13.0-45-generic #50-Ubuntu 
CPU model name  : Intel(R) Xeon(R) CPU @ 2.00GHz
cpu cores       : 16
Hardware threads : 32
Max threads per process : 8
MemTotal:       29631356 kB
L3 cache size   : 56320 KB
CPU instructions width : popcnt:4, sse4.2:256, avx:512, avx2:1024
Using instructions : AVX-2, popcnt

Compiler : gcc 7.2.0
Compiled on : Jul 24 2018 02:36:06
Executed on : 7/27/2018 7:9:14
Size of complex : 8 B

Verbosity : 3
Circuit file : inst_7x7_49_0
Circuit type : Google
Qubits : 49  Gates : 1516 (504 two q gates) Cycles : 50
Qubit grid 7x7 with 2 blocks 
     0^   1^   2^   3^   4^   5^   6^  
     7^   8^   9^  10^  11^  12^  13^  
    14^  15^  16^  17^  18^  19^  20^  
    21^  22^  23^  24^  25^  26^  27^  
    28_  29_  30_  31_  32_  33_  34_  
    35_  36_  37_  38_  39_  40_  41_  
    42_  43_  44_  45_  46_  47_  48_  
 Cross-gates : 7

Simulation type : sum of tensor products / single cut
Cut : horizontal 28q + 21q (42 xCZ gates)
Requested end-to-end circuit fidelity: 2.38e-07
Approximation type : pruned xCZ branches
Simulating xCZ gates : using projection-based branches
xCZ path breakdown : 35p + 4r + 3b (cycle breakdown : 45p + 4r + 1b)
Low-value qubits : 14 q, 11 q
Requested num amps : 1000000

Multi-process simulation : 
     8192 processes (8 threads each) over 512 nodes in 4 batches per node 
     State representation size : 2.02 GiB 
     Peak memory : 12410.88 GiB (24.24 GiB per node)
     Layers simulated : H (161), CZ & T (215), X & Y & H (71)
     Layers breakdown : 24p + 64r + 128b
     Batch stats :
          Avg user time : 5475.45 s 
          Wallclock : 1074.158 s (avg), 1536.13 s (max)
          Avg CPU utilization : 582.699% (72.837% per thread)
          Avg resident size : 5.985 GiB
          Avg page faults : 0.63 (major), 9500721.534 (minor)
     Billable runtime : 2.185e+02 hrs (2.185e-04 hrs per amp)
     Avg zero count 
          1st Checkpoint : A = 0 (0.0%), B = 0 (0.0%)
          2nd Checkpoint : A = 16777216 (6.25%), B = 0 (0.0%)

amp[3]          = -1.921780e-11-2.707243e-12j
amp[1/4]        = 1.118916e-11-1.348977e-11j
amp[1/2]        = -4.073272e-12-9.529898e-13j
amp[3/4]        = -1.650529e-11+9.459437e-12j
amp[-3]         = 2.059435e-11-1.063607e-11j

Avg runtime (265.755 s total) per process by category 
     Initial H (49)                          : 0.13483 s     = 0.051%
     Last H (25)                             : 97.202 s      = 36.576%
     xCZ (42)                                : 21.841 s      = 8.219%
     CZ & T (811), Low X & Y (270) & H (18)  : 29.279 s      = 11.017%
     Single X (8) & Y (5)                    : 5.889 s       = 2.216%
     High X & Y (282) & H (6)                : 33.962 s      = 12.779%
     Rescaling passes (2)                    : 0.374 s       = 0.141%
     Copying (127)                           : 29.357 s      = 11.047%
     Storing amps                            : 35.275 s      = 13.274%
                                                               ----------
     Total                                                     95.319%

Avg time per gate : 0.00137 s
Simulation per process runtime breakdown : 
     Prefix                  : 119.984 s = 45.148%
     Branches                : 171.701 s = 64.609%
     Memory mapped I/O       : 0.00753 s = 0.00283%

\end{verbatim}

\section{ 56-qubit depth $1+40+1$ simulation with fidelity 0.51\%}
This simulation uses the circuit \verb+inst_7x8_41_0+ publicly available from \cite{new_benchmarks}.

\begin{verbatim}

(C) 2017, 2018  Regents of the University of Michigan
Rollright ver 2.2 - a quantum circuit simulator
Igor L. Markov and Aneeqa Fatima

Hostname : Linux boixo-rr-hs32-28 4.13.0-43-generic #48-Ubuntu
CPU model name	: Intel(R) Xeon(R) CPU @ 2.00GHz
CPU cores	: 16
Hardware threads : 32
Max threads per process : 8
MemTotal:       123768660 kB
L3 cache size	: 56320 KB
CPU instructions width : popcnt:4, sse4.2:256, avx:512, avx2:1024
Using instructions : AVX-2, popcnt

Compiler : gcc 7.2.0
Compiled on : Jun  8 2018 07:28:48
Executed on : 6/20/2018 18:18:13
Size of complex : 8 B

Verbosity : 3
Circuit file : inst_7x8_41_0
Circuit type : Google
Qubits : 56  Gates : 1466  Cycles : 42
Qubit grid 7x8 with 2 blocks
     0^   1^   2^   3^   4_   5_   6_   7_
     8^   9^  10^  11^  12_  13_  14_  15_
    16^  17^  18^  19^  20_  21_  22_  23_
    24^  25^  26^  27^  28_  29_  30_  31_
    32^  33^  34^  35^  36_  37_  38_  39_
    40^  41^  42^  43^  44_  45_  46_  47_
    48^  49^  50^  51^  52_  53_  54_  55_
 Cross-gates : 7

Simulation type : sum of tensor products / single cut
Cut : vertical 28q + 28q (35 xCZ gates)
Requested end-to-end circuit fidelity: 0.0051
Approximation type : pruned xCZ branches
Simulating xCZ gates : using projection-based branches
xCZ path breakdown : 28p + 3r + 4b (cycle breakdown : 37p + 4r + 1b)
Low-value qubits : 14 q, 14 q
Requested num amps : 1000000

Multi-process simulation :
     1369569 processes (8 threads each) over 625 nodes in 4 batches per node
     State representation size : 4 GiB
     Peak memory : 30000.0 GiB (48.0 GiB per node)
     Layers simulated : H (145), CZ & T (179), X & Y & H (179)
     Layers breakdown : 20p + 32r + 128b
     Batch stats :
          Avg user time : 418.77 hrs
          Wallclock : 87.592 hrs (avg), 140.744 hrs (max)
          Avg CPU utilization : 483.757% (60.47% per thread)
          Avg resident size : 11.774 GiB
          Avg page faults : 0.002 (major), 10508777.493 (minor)
     Billable runtime : 8.796e+04 hrs (8.796e-02 hrs per amp)

amp[3]  	= 2.911096e-10-1.236132e-11j
amp[1/4]	= 4.302013e-13+1.527466e-10j
amp[1/2]	= -7.879150e-11-2.773058e-10j
amp[3/4]	= 2.147728e-10-6.191063e-11j
amp[-3] 	= -3.276638e-11+5.329280e-11j

Avg runtime (570.988 s total) per process by category
     Initial H (56)                          : 0.22702 s    = 0.04%
     Last H (28)                             : 51.494 s     = 9.018%
     xCZ (35)                                : 41.237 s     = 7.222%
     CZ & T (781), Low X & Y (258) & H (24)  : 112.343 s    = 19.675%
     Single X (8) & Y (16)                   : 30.526 s     = 5.346%
     High X & Y (256) & H (4)                : 163.514 s    = 28.637%
     Rescaling passes (10)                   : 2.547 s      = 0.446%
     Copying (136)	                          : 58.704 s     = 10.281%
     Storing amps                            : 99.107 s     = 17.357%
                                               ----------
     Total                                     98.023%

Avg time per gate : 0.003043 s
Simulation per process runtime breakdown :
     Prefix               : 106.885 s = 18.719%
     Branches             : 461.465 s = 80.819%
     Memory mapped I/O    : 0.00913 s = 0.0016%

\end{verbatim}

\end{document}